\documentclass[journal]{IEEEtran}
\newcommand{\bl}{\textcolor{black}}

\usepackage{graphicx,epsfig,amssymb,amsmath,url,latexsym,stfloats,array,bm}
\usepackage[tight,footnotesize]{subfigure}
\usepackage{makecell}
\usepackage{mathrsfs}
\usepackage{amsthm}
\usepackage{longtable}
\usepackage{amsfonts}
\usepackage{caption}
\usepackage{graphicx}
\usepackage{multirow}
\usepackage{longtable}
\usepackage{float}
\usepackage{afterpage}
\usepackage{cite}
\usepackage{subeqnarray}
\usepackage{epstopdf}
\usepackage{empheq}
\usepackage{latexsym}
\usepackage{CJK}
\usepackage{color, soul}
\usepackage{framed}
\usepackage{lipsum}
\usepackage{color}
\usepackage{stfloats}
\usepackage{setspace}
\usepackage{hyperref}
\definecolor{shadecolor}{rgb}{1,0,0}

\usepackage{algorithmic}
\usepackage[linesnumbered,ruled]{algorithm2e}
\usepackage{cuted}

\usepackage{url}

\begin{document}
\title{Toward Immersive Communications in 6G}

\author{ Xuemin~(Sherman)~Shen,~\IEEEmembership{Fellow,~IEEE,}
            Jie~Gao,~\IEEEmembership{Senior~Member,~IEEE,}
            Mushu~Li,~\IEEEmembership{Member,~IEEE,} Conghao~Zhou,~\IEEEmembership{Member,~IEEE,} Shisheng~Hu,~\IEEEmembership{Student~Member,~IEEE,}
            Mingcheng~He,~\IEEEmembership{Student~Member,~IEEE,}
             and
            Weihua Zhuang,~\IEEEmembership{Fellow,~IEEE}

\thanks{X.~Shen, C.~Zhou, S.~Hu, M.~He, and W.~Zhuang are with the Department of Electrical and Computer Engineering, University of Waterloo, Waterloo, ON, N2L 3G1, Canada (e-mail:\{sshen, c89zhou, wzhuang\}@uwaterloo.ca).}
\thanks{J.~Gao is with the School of Information Technology, Carleton University, Ottawa, ON, K1S 5B6, Canada (email:~jie.gao6@carleton.ca).}
\thanks{M.~Li. is with the Department of Electrical, Computer, and Biomedical Engineering, Toronto Metropolitan University, Toronto, ON, M5B 2K3, Canada (e-mail:~mushu1.li@ryerson.ca)}
}   
\maketitle

\begin{abstract} 
The sixth generation (6G) networks are expected to enable immersive communications and bridge the physical and the virtual worlds. Integrating extended reality, holography, and haptics, immersive communications will revolutionize how people work, entertain, and communicate by enabling lifelike interactions. However, the unprecedented demand for data transmission rate and the stringent requirements on latency and reliability create challenges for 6G networks to support immersive communications. In this survey article, we present the prospect of immersive communications and investigate emerging solutions to the corresponding challenges for 6G. First, we introduce use cases of immersive communications, in the fields of entertainment, education, and healthcare. Second, we present the concepts of immersive communications, including extended reality, haptic communication, and holographic communication, their basic implementation procedures, and their requirements on networks in terms of transmission rate, latency, and reliability. Third, we summarize the potential solutions to addressing the challenges from the aspects of communication, computing, and networking. Finally, we discuss future research directions and conclude this study.
\end{abstract}


\section{Introduction}\label{s:Intro}

Ever since its birth, communication technology has been a symbol of the modernization of human society, and the evolution of communication technology has accompanied the advance of civilization.
The commercialization of electrical telegraph and telephone during the second industrial revolution boosted globalization by facilitating finance and trade overseas~\cite{wenzlhuemer2013connecting}.  The debut of vehicle-mounted mobile radio systems (``car phones'') and the analog first generation (1G) mobile telecommunication systems from the 1950s to 1980s enabled voice calls on the go~\cite{8226757}. The second generation (2G) mobile communication systems, which introduced roaming and preliminary data services in the form of text messages, emerged amidst and as a part of the third industrial revolution (i.e., the digital revolution)~\cite{billstrom2006fifty}. Then, the next two decades witnessed the proliferation of mobile Internet and mobile multimedia services brought by the third and fourth generation  (3G and 4G)  mobile communication technology, which revolutionized how people communicate and changed the world. Nowadays, the fifth generation (5G) mobile communication systems are reshaping industries by facilitating the fourth industrial revolution (i.e., Industry 4.0) towards smart inter-connectivity and automation~\cite{9272626}.  

Accustomed to the convenience brought by the latest communication technologies, many people may not realize that ordinary daily activities such as video calls or zoom meetings were nothing more than science fiction merely three decades ago. Indeed, from the so-called ``telephot" in the pioneering novel ``Ralph 124C 41+" to the video call scene in the classic movie ``Back to the Future", the simultaneous transmission of live image and sound was considered as a ``technology of the future'' in the most part of the twentieth century~\cite{gooday2005electrical, fowler1986back}. When the fantasy of the past has become a reality, a question that naturally arises is: what will be the next revolutionary form of communications, potentially in the era of the sixth generation (6G)? Fortunately, we may again find clues in science fiction, with examples ranging from the famous scene of Princess Leia's three-dimensional (3D) holographic message in ``Star Wars"~\cite{4752917} to the virtual world ``OASIS'' in the metaverse presented in the recent film ``Ready Player One"~\cite{SPARKES202118}. The fact that such scenes created a long-lasting influence on a vast audience reflected people's desire for more lifelike, immersive, and interactive communications~\cite{xu2022full}.   

Unfolding exactly as depicted in science fiction or not, immersive communications  will come to reality and shift the current communication paradigm in three aspects. First, rather than two-dimensional (2D) images displayed on a flat screen, immersive communications will deliver 3D images with parallax information. Second, in addition to audiovisual information, immersive communications will involve haptic information. Third, the pursuit of immersive experiences will further blur the boundary between the physical and the virtual worlds, allowing new forms of interactions across the two worlds. These paradigm shifts can significantly enrich communication experiences of users and enable a plethora of new use cases such as 3D telepresence~\cite{9523831},  ultra-realistic online interactive sports~\cite{NxGAllianceUseCases}, and immersive learning in education~\cite{9177354}, to name a few. In particular, immersive communications can also enable human-machine collaboration in industrial environments and propel the next industrial revolution, i.e., industrial 5.0~\cite{MADDIKUNTA2022100257, leng2022industry}. As a result, immersive communications are expected to have a profound impact on the landscape of communication industries and impact how people study, work, and entertain in the years to come. 

Motivated by the potentials of immersive communications, scientists and engineers over the world have been working on the development of related technologies, products, and platforms. Significant progress has been made in recent years, including but not limited to advancements in sensor systems and data capture techniques~\cite{10.1145/3517031.3529616,8858052}, data processing and computing frameworks~\cite{9721178, 9772902, qian2022remote}, and rendering and display devices~\cite{hirayama2019volumetric, xiong2021augmented, 9000715}.  Some component development of immersive communications is progressing faster than others, leading to the establishment of testbeds, prototypes, or even commercial products. Virtual reality (VR), as an example, has gained popularity, especially in the gaming industry~\cite{jung2020augmented}. Devices such as VR headsets and haptic glove development kits are available in the market~\cite{chen2022simulation, kugler2021state}, while researchers are building testbeds for extended reality (XR)~\cite{9741292} and human-machine interaction with haptic feedback~\cite{9063407}. An example of recent development in immersive communications is the VirtualCube system, a 3D video conference system capable of synthesizing remote and local participants so that they appear in the same environment~\cite{9714050}. In addition, a research team in Germany is exploring VR-based full-body avatars for training police forces while evaluating their stress level and response to threats~\cite{9706353}.

As the aforementioned progress and efforts are paving the way for realizing immersive communications, advancements in communication and networking technologies will be indispensable. Despite the advent of 5G systems and the accompanying advancements in network capabilities, there are still many challenges to achieving immersive communications in various aspects of communications, networking, and computing. The data rate required to transmit live 3D images can be so high, e.g., on the level of terabits per second (Tbps), that even 5G cannot support it, especially for high-resolution and $360^{\circ}$ videos. The required end-to-end delay for delivering haptic information can be as low as a few milliseconds for a satisfactory user experience~\cite{maier2019towards, sim2021low}. The synchronization of data streams from multiple cameras or sensors and that of audiovisual and haptic information in data transmission also create new challenges. The storing and processing of massive data for immersive communications demand new architectures and techniques for caching and computing~\cite{9156256, taleb2021extremely, 9537928}. Moreover,  artificial intelligence (AI) is necessary both for supporting applications such as human-machine collaboration and user viewpoint/gesture prediction, and for orchestrating network resources to satisfy the demanding requirements of immersive communications~\cite{tataria20216g, zawish2022ai, 8423605}. Since the realization of immersive communications can require integrated support for enhanced mobile broadband (eMBB) and
ultra-reliable low-latency communications (URLLC)~\cite{pang2022new}, which is beyond the capability of 5G, researchers look forward to breakthroughs in immersive communications in the era of 6G. Targeting 2030 for large scale 6G deployments, the 3rd Generation Partnership Project (3GPP) plans to start 6G studies in 2024 and complete its first 6G standard in 2028~\cite{Ericsson5GA6G2022}, while the International Telecommunication Union (ITU)'s ``IMT for 2030 and beyond'' timeline aims at completing IMT-2030 specifications in 2029 to 2030~\cite{9815809}.

Recognizing the importance of immersive communications, the research community in communications, networking, and computer science is expanding its effort in this field. Several recent review and survey articles can be found in the literature, among which some present state of the art in immersive communications, while others envision the next steps. Most of these articles focus on a specific aspect, such as supporting 360{$^\circ$}/holographic video streaming~\cite{9133103, 9915358}, evaluating the immersive experience of users~\cite{9679801}, analyzing the effects of user motions on network performance in XR~\cite{9921198}, enabling the use case of Metaverse~\cite{xu2022full,9880528, 9806418}, or facilitating distributed implementation of VR~\cite{9722829}. Different from the above works, we present a comprehensive survey of immersive communications in this article. With a focus on the communication, networking, and computing perspectives, 
we review a large number of publications, especially the latest works in  communications, networking, and computer science to present the representative use cases, the recent developments, the technical challenges, and the potential solutions related to immersive communications in the era of 6G communications. In specific, we focus on immersive communications by looking into its three main forms, i.e., XR, haptic communication, and holographic communication in the remainder of this article. Section~\ref{s:UCs} introduces representative use cases of immersive communications to illustrate its promising prospect. Section~\ref{s:Comps} presents the concepts, basic implementation procedures, and requirements of XR, haptic communication, and holographic communication to paint an overall picture of immersive communications. Section~\ref{s:Solus} focuses on the challenges and the state-of-the-art solutions towards realizing each of the three forms of immersive communications. Section~\ref{s:FDs} discusses some open issues regarding immersive communications in 6G, and Section~\ref{s:Con} concludes this article. A list of the main
acronyms used in this article is given in Table~\ref{acronym}.
\begin{table*}
    \centering
    \captionsetup{justification=centering,singlelinecheck=false}
    \caption{List of main acronyms}\label{acronym}
    \begin{tabular}{l|l||l|l}
        \hline\hline
        1G - 6G & First Generation - Sixth Generation & 2D & Two-dimensional \\
        3D & Three-dimensional & 3GPP & 3rd Generation Partnership Project\\
        AI & Artificial Intelligence & AR & Augmented Reality \\
        D2D & Device-to-device & DetNet & Deterministic Networking \\
        DoF & Degree of Freedom & FoV & Field-of-view\\
        HI & Haptic Interface & IMU & Inertial Measurement Unit \\
        IRS & Intelligent Reflecting Surface & JND & Just-noticeable Difference \\
        LDPC & Low-density Parity-check & LFV & Light-field Video \\
        LIDAR & Light Detection and Ranging & LoS & Line-of-sight\\
        LSTM & Long Short-term Memory & MEC & Mobile Edge Computing\\
        MIMO & Multiple-input and Multiple-out & MR & Mixed Reality \\
        MSE & Mean Squared Error & MTP & Motion-to-photon\\
        MVC & Multi-view Coding & NOMA & Non-orthogonal Multiple Access\\
        O-RAN & Open Radio Access Network & PoV & Persistence of Vision\\
        QoE & Quality of Experience & QoS & Quality of Service\\
        THz & Terahertz & TSN & Time-sensitive Networking\\
        VR & Virtual Reality & XR & Extended Reality\\
        \hline\hline
    \end{tabular}
\end{table*}

\section{Use Cases}\label{s:UCs}
There are many potential use cases for immersive communications, relating to both commercial and enterprise scenarios and ranging from gaming to industrial control. In this section, we detail four representative use cases to illustrate the promising prospect of immersive communications. A list of representative use cases is given in Table~\ref{Table1}.  

\subsection{Immersive Gaming and Entertainment}

{XR provides the ultimate gaming and entertainment experience by presenting convincing gaming environments through XR devices such as VR headsets or smartphones. Players can interact with each other without feeling a barrier between the virtual and the physical worlds \cite{Bastug}.  XR devices display the virtual world of the game to players and capture their actions such as eye movements to allow them to interact with the virtual world \cite{Elbamby}. With the success of advanced XR gaming consoles and headsets, e.g., Oculus and PlayStation VR, as well as games and platforms, e.g., Pokemon Go and Roblox, game developers are striving to offer more flexible XR experiences with wireless XR devices \cite{Maimone}. Through wireless XR devices, players can interact freely with other players or virtual objects, e.g., in XR sporting~\cite{Kim}. Furthermore,  haptic communication devices can be combined with XR to significantly enhance the immersive gaming experience. Transducer arrays, which can be attached to XR devices, can capture haptic data from players. As a result, XR devices can fuse haptic information into the virtual world and provide haptic feedback to players by mapping motions in the game to players' sensations. Players can use haptic devices, such as gloves, to control objects  in the game \cite{hashimoto2006group} or synchronize their sensations with other players \cite{mauve2000consistency}. }

\subsection{Telesurgery}

In telesurgery, surgeons remotely manipulate robotic arms to operate on patients by utilizing control panels and low-latency display of the surgical scenes. Telesurgery is beneficial in removing the barrier of distance among surgeons and patients, tackling the scarcity of surgeons in remote or difficult-to-reach areas such as countryside, battlefields and spacecraft, and facilitating the collaboration of surgeons at different locations \cite{choi2018telesurgery, mohan2021telesurgery}. The assistance of robotic arms can enhance the performance of surgeries by detecting and canceling out the physiological tremors of surgeons’ hand motions \cite{kumar2020real}, performing delicate surgical operations and minimizing the surgical incision areas for reducing blood loss and incision-related complications \cite{diana2015robotic}. To guarantee the performance of surgeons, the display of surgical scenes to them should be highly precise and informative. To this end, 3D video of the surgical scenes with depth information, can be displayed to the surgeons, e.g., by using passive polarized glasses, and an eye-tracking mechanism can be used to quickly center the area where the surgeon is viewing in the visual display \cite{doi:10.3109/13645706.2014.1003945}. In addition, augmented reality (AR) can be leveraged to overlay medical images such as ultrasound images and computed tomography (CT) images onto the video of surgical scenes \cite{10.1117/1.JMI.3.4.045001}. Besides visual information, haptic information in the surgeries, such as the texture of tissues and the tension in tying surgical sutures, can be captured by the haptic devices on the robotic arms and then transmitted to and reproduced by the haptic devices at the surgeons' side \cite{patel2022haptic,el2020review}.

\subsection{Immersive Learning}

Immersive learning integrates emerging technologies, including XR and haptic technologies, into teaching to provide students or trainees an interactive and engaging learning experience~\cite{affan2021haptic,laamarti2014overview}. During the recent COVID-19 pandemic, traditional methods of teaching, e.g., online courses, encountered the problem of engaging students in the learning process~\cite{jumreornvong2020telemedicine, fitzek2021tactile}.
To this end, immersive learning, as a potential solution to boost student engagement, is receiving increasing attention, especially from primary and secondary schools. With immersive learning, avatars of students and teachers can be created in the virtual world~\cite{gupta2019tactile}, and each student is allowed to interact with the avatars of teachers and other students via the senses of sight, hearing, and touch. Such interactions can keep students' attention in learning process. Immersive learning 
is categorized as either asynchronous or synchronous. Training some skills, such as sports skills and cooperative tele-operation skills for industrial robots, requires real-time interactions, which can encourage active participation in the learning process~\cite{kaluschke2021shared,lee2021real}. Utilizing XR, haptic communication, and holography communication technologies, teachers can check whether the moves and actions of their students are correct and provide immediate corrections if not, regardless of their physical distance from each other. For the skills that do not need real-time interactions, information regarding teachers' positions, velocities, and applied forces can be recorded and displayed to students via XR and haptic devices asynchronously~\cite{tan2020methodology}. Such ``record-and-replay'' strategy can allow a much larger number of students to learn at their own pace, despite the absence of real-time interactions~\cite{steinbach2018haptic,yokokohji1996you,yokokohji1996toward}.

\subsection{Holographic Teleconference}
Teleconference is a convenient choice for users to remotely collaborate with each other. 
In the current video teleconferencing, remote participants can only be displayed on flat screens, which results in a very different perception in a virtual conference from that in an on-site conference. 
In order to provide an immersive experience in teleconferences, holographic teleconferences depict realistic 3D presence for people by projecting 3D images of remote participants as holograms \cite{jiang2021road, 9714050,zhou2022intelligence, siemonsma2022holokinect}.
Specifically, when a remote participant joins the holographic teleconference, 3D visual information and the corresponding audio information of the participant can be captured by multiple sensors, transmitted, and then reconstructed as a hologram on the side of other participants to provide 3D audiovisual information for interactions among participants \cite{strinati20196g}.
In this case, holographic teleconference can reduce the impact on participants of the separation between the virtual and the physical worlds.
In addition to the audio and video information, participants in a holographic teleconference are able to obtain haptic information from others to achieve an immersive experience with the sense of physical contacts \cite{tataria20216g}.
For example, a participant with haptic sensors can sense a handshake with others, thereby enabling an immersive experience similar to in-person interactions.

\subsection{Metaverse}
A metaverse provides fully immersive and self-sustaining virtual spaces that merge the physical and digital worlds \cite{wang2022survey}. In the metaverse, users can have avatars as digital representations in simulated or imaginary environments, such as games and virtual cities. Through XR devices including phones or laptops, users can interact with digital avatars, other digital objects, and virtual environments. Metaverses require the synchronization between the physical and the digital worlds through two main information flows. One of them is from the physical world to digital worlds, in which sensors and actuators capture user activity so that the behaviors of a user in the physical world are reflected via their avatars in a digital world. The other is from digital worlds to the physical world, including the interactions among avatars, other digital objects, and metaverse services in the virtual environments. As a result of advanced networking technologies, big data analysis, blockchain, and AI, metaverses are expected to provide human-centric content for users to enable immersive social experiences \cite{metaverse1}, online collaborations \cite{suzuki2020virtual}, etc.   

\begin{table}[htbp]
    \caption{Representative use cases of immersive communications}\label{Table1}
    \centering
    \renewcommand{\arraystretch}{1.5}%
    \begin{tabular}{l|p{0.48\linewidth}}
        \hline\hline
            \textbf{Use Cases} & \textbf{References}\\ 
        \hline\hline
            Gaming  & \cite{carroll2021effect, 8726071,rega2022free} \\ 
            \hline
            \multirow{2}*{E-learning} & \cite{kavanagh2017systematic,harvey2021comparison, ahmad2021edugram} \\ 
                                      & \cite{makransky2021cognitive} \\
            \hline
            Teleconference & \cite{siemonsma2022holokinect, 9714050} \\ 
            \hline
            Tele-operation & \cite{lee2021real,choi2018telesurgery}  \\ 
            \hline
            E-heath & \bl{\cite{jumreornvong2020telemedicine,velana2022advances}} \\ 
            \hline
            E-commerce & \cite{ornati2022fashion,speicher2017vrshop}  \\ 
            \hline
            Smart home & \cite{zhu2020haptic,eckstein2019smart} \\ 
            \hline
            Manufacturing & \cite{aijaz2018tactile,lipton2017baxter}  \\ 
            \hline
            Tourism and travel & \bl{\cite{chun2022traveling,han2018identifying}}\\ 
            \hline
            \bl{Metaverse} & \bl{\cite{9880528,wang2022survey,xu2022full,yang2022fusing}}\\
        \hline\hline
    \end{tabular}  
\end{table}

\section{Immersive Communications: Concepts and Requirements}\label{s:Comps}

The use cases for immersive communications and their potential importance in 6G are intuitive. Understanding immersive communications beyond the use cases, however, requires answers to the question ``what are immersive communications?".  Since the research of immersive communications is in an early stage, there is no commonly-agreed definition yet. 

\begin{figure*}[htbp]
		\centering
	  	\includegraphics[width=0.6\textwidth]{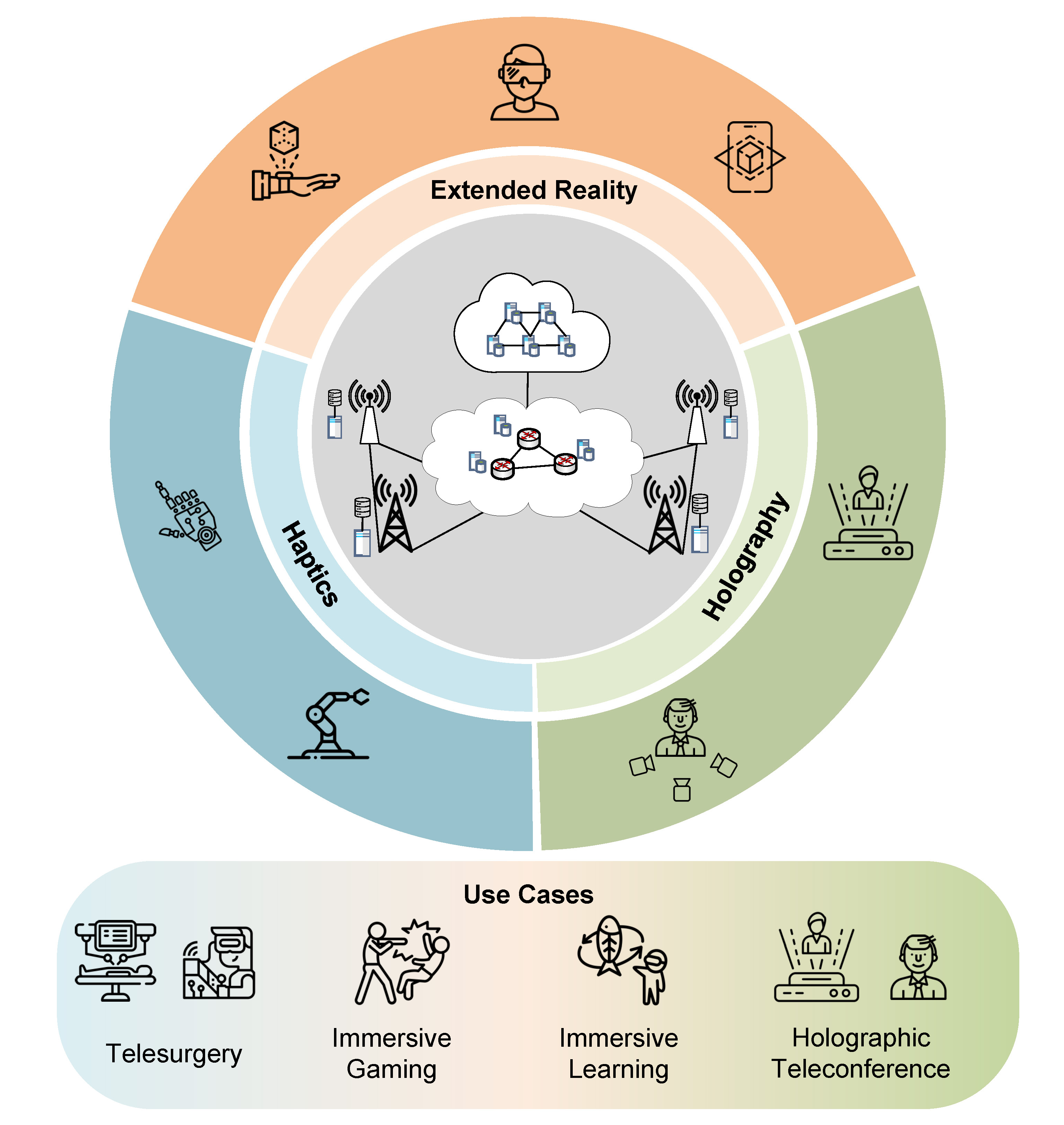}
	  	\caption{Main forms and exemplary use cases of immersive communications.}\label{fig:1}
\end{figure*}

\begin{figure*}[htbp]
		\centering
	  	\includegraphics[width=0.6\textwidth]{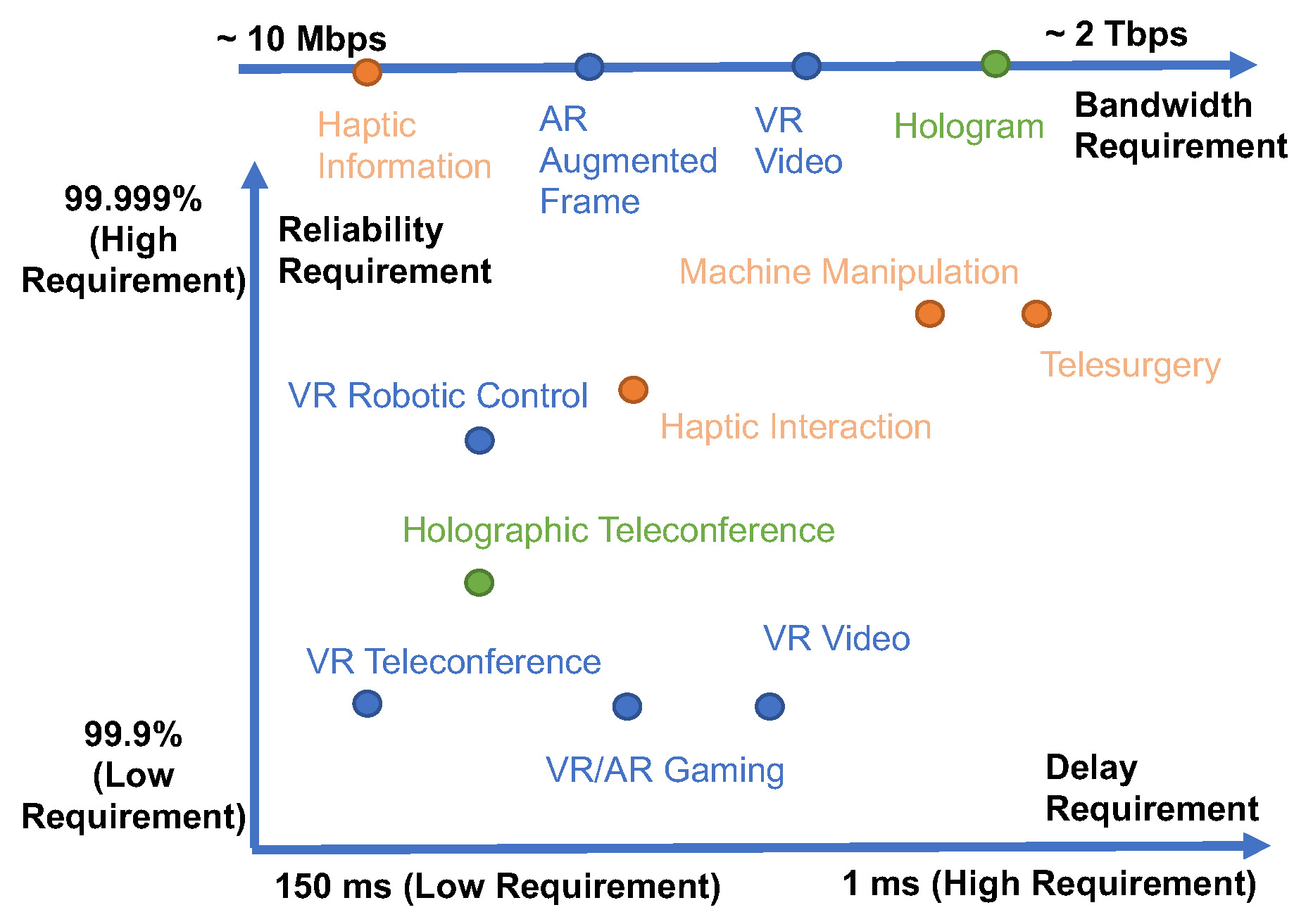}
	  	\caption{Requirements of representative immersive communications use cases: an illustration.}\label{fig:new}
\end{figure*}

We consider immersive communications as a communication paradigm along with the supporting technologies that allow users to have lifelike experiences in the physical world, the virtual world, or both, with interactions via 3D audiovisual and/or haptic information exchange. In this section, we focus on the three main forms of immersive communications as illustrated in Fig.~\ref{fig:1}, i.e., XR, haptic communication, and holographic communication.\footnote{Note that the three forms may co-exist since a use case may involve more than one form, and additional forms of immersive communications may exist or emerge.} Via introducing the concept, basic implementation procedure, and the network requirements for each of the three forms, we aim to sketch an overall picture of immersive communications. The requirements of representative immersive communications use cases  are illustrated in Fig.~\ref{fig:new} and also summarized in Table~\ref{Table2}.

\subsection{Extended Reality}\label{sss:XR}
In this subsection, we introduce the concept of XR and investigate two respective XR technologies: VR and AR. Then, we examine their implementation procedures and service requirements for 6G. 
\subsubsection{Concept}
{XR covers a range of technologies, including VR, AR, mixed reality (MR), and everything in between \cite{Hu}. In general, XR combines the physical and virtual worlds through extensive video processing and data fusion. Using XR devices, users can interact with virtual avatars and access XR content. Under the umbrella of XR, a variety of technologies are defined depending on the level of virtuality. Two representative technologies in XR are AR and VR.  With the lowest level of virtuality, AR focuses on constructing artificial objects according to the objects (e.g., buildings, faces, or vehicles) residing in the physical world and enabling users to interact with them. Conversely, with the highest level of virtuality, VR creates an entirely artificial scenery and allows users to interact with the objects in a completely artificial environment generated by the headsets.  In MR, the concepts of VR and AR can be combined to create different levels of virtuality.
In spite of the variety of XR technologies, the methods to provide immersive experiences to users are similar, which combine sensory data with virtual environments to produce artificial sceneries, from either the physical or virtual worlds, using headsets or portable display devices.}

{The first VR flight simulator was developed in 1970s to train pilots for flights without exposing them to risks of flying \cite{earnshaw2014virtual}. In the early stage, VR headsets were cumbersome, and processing VR content required large supercomputers. Nowadays, VR technologies have gained momentum due to recent advances in computing and display technologies. The headsets, such as Oculus head-mounted displays and HTC Vive, are affordable and can  support ultra-high resolutions (3840$\times$2160 in Pimax 8K) and refresh rates (up to 120 Hz) \cite{Hu}.

Most VR content is processed and rendered by user devices. Rendering content with a high level of virtuality requires extensive computing power. For a VR headset, a console is required to supply additional computing power to the headset, while a wired connection restricts the user to a workstation. Therefore, wireless VR is the primary focus of VR research now \cite{Elbamby2}. 
In addition, multi-sensory XR, as another future vision of XR, integrates human senses and perception, including visual, auditory, olfactory, and tactile into XR content, enabling a truly immersive experience. This requires the confluence of multiple disciplines, including AI, computer vision, biology, ultra-low-latency networking, etc., while linking the real and virtual worlds \cite{Fenghe, Yuting}.
}  

\subsubsection{Basic Implementation Procedure}

{While XR comprises several technologies with different levels of virtuality, its implementation procedures can be summarized into three steps: content transmission, rendering, and feedback collection. For each of the above three steps, communication networks can play an important role.}

\begin{figure*}[t]
		\centering
	  	\includegraphics[width=0.8\textwidth]{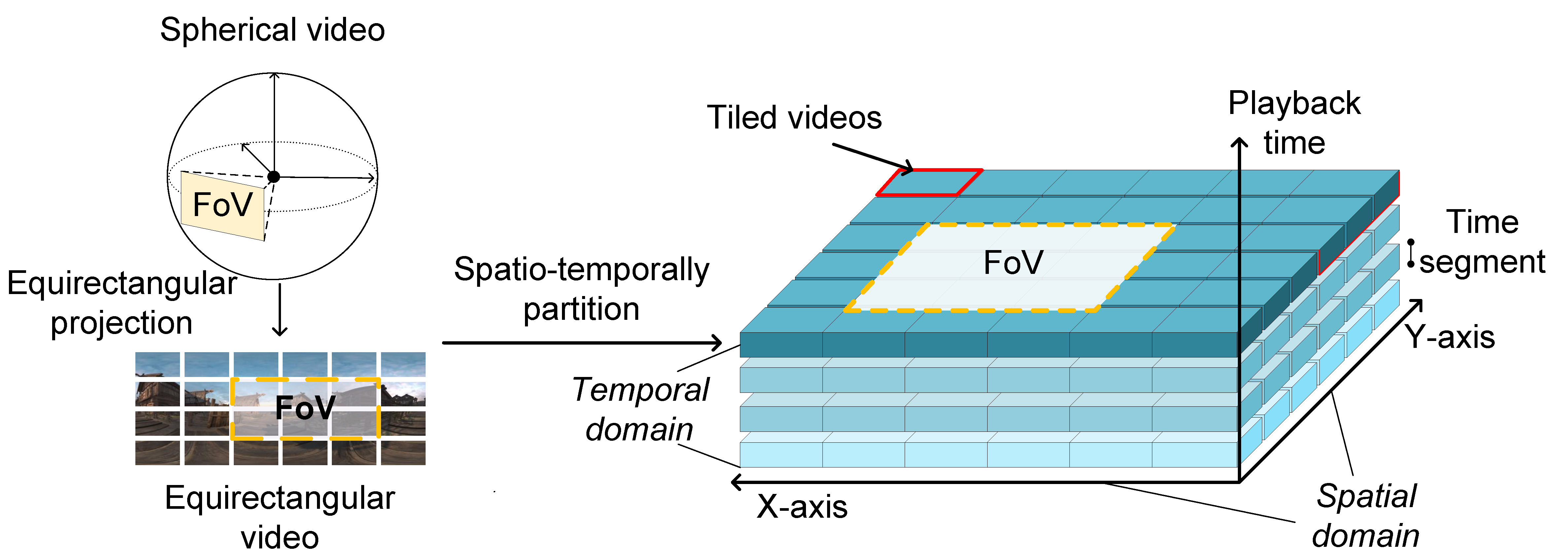}
	  	\caption{VR video projection and partition.}\label{fig:XR_1}
    
\end{figure*}

{In the step of content transmission, VR content generated by VR content providers is transmitted from content servers and VR devices.  VR devices play 360\textdegree~ spherical videos, which can be mapped to equirectangular videos. During playing VR content, these equirectangular videos are mapped onto a sphere, in which the user is situated at the center, to provide a 3D stereoscopic experience. The key feature of VR video is the ultra-high spatial resolution. A VR video has a resolution of up to 12K (11,520 $\times$ 6,480), while the conventional video normally has a resolution of 4K or less. Transmitting full equirectangular videos from content servers requires an ultra-high data rate. Thus, tile-based transmission is usually adopted in VR video delivery. As shown in Fig. \ref{fig:XR_1},  a content server can divide equirectangular videos spatio-temporally into video chunks, i.e., tiled videos, and only the tiled videos within a user's field-of-view (FoV) is delivered \cite{Praveen, Jangwoo}. In this way, VR content can be delivered in a significantly reduced data size. However, the tile-based solution requires VR headsets to detect and estimate user viewpoints to determine the region of FoV. Content servers should select which tiled videos to be delivered to users based on both the user's current viewpoint and network conditions \cite{Zare}. In terms of AR, AR devices generate raw content by the sensors at the local devices, such as cameras in smartphones \cite{Ren}. In contrast to VR devices, which download content from a content server, AR devices can upload raw content to the server for further processing. Specifically, raw videos captured by AR devices are clipped into frames with a specific image format, and those frames can be offloaded to the server. The processed content is then delivered to and played on the AR devices. }

\begin{figure*}[htbp]
		\centering
	  	\includegraphics[width=0.8\textwidth]{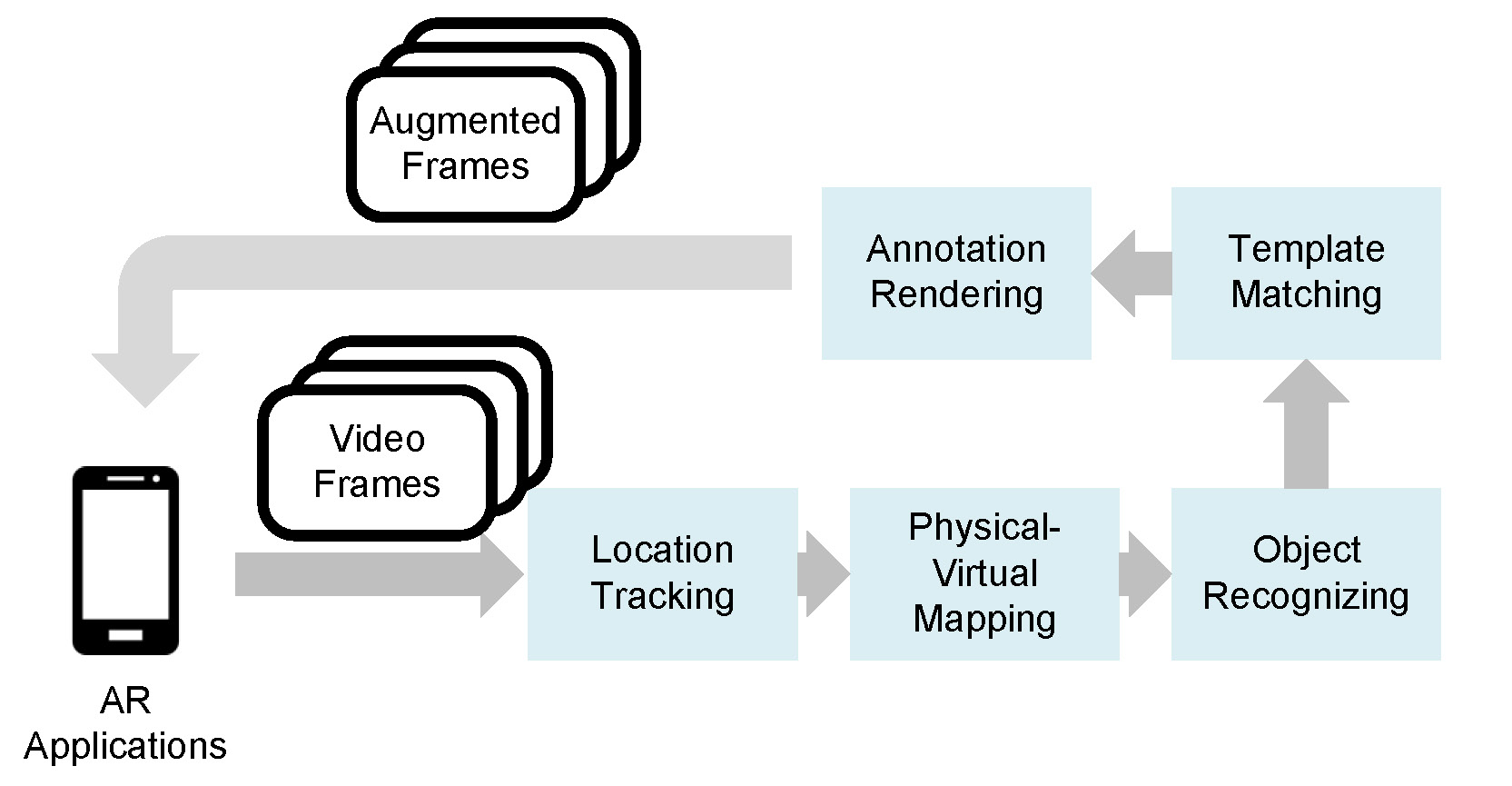}
	  	\caption{ Content processing and rendering for AR applications.}\label{fig:XR_2}
\end{figure*}
{In the step of content rendering, tiled VR videos transmitted to VR devices are stitched together, and computing resources are required to project 2D stereoscopic videos to 3D stereoscopic videos, i.e., generating two different videos for the left and right eyes respectively. This content rendering step can be performed on VR devices once all the required content has been received. In addition, due to the limited computing capability of VR devices, the workload of content rendering can be offloaded to adjacent edge servers enabled by mobile edge computing (MEC) \cite{Jianmei,Dang,sukhmani2018edge}. Content processing and rendering are more complex in AR than in VR, where AR processing procedures are shown in Fig. \ref{fig:XR_2}. Once the raw AR content, i.e., video frames, is captured by an AR device, a location tracking step determines the device's location and position according to the captured frames. Then, a mapping step establishes a virtual coordinate of the environment based on the result of the tracker, and an object recognizing step detects the objects to process in the video frames \cite{Xiuquan, Jinke}. Based on the identified objects, the augmented data is retrieved from the local cache or network servers and attached to the frames accordingly. Specifically, a template matching step attaches the augmented data to the frames, and an annotation rendering step renders the processed frames at AR devices. The computing workload for conducting the above functions can be fully or partially offloaded from AR devices to network servers to minimize computing latency or improve energy efficiency at AR devices. }

After receiving and playing XR content, XR devices collect user feedback to select the content to deliver next. VR and AR devices have similar methods for feedback collection, with sensors or cameras attached to the devices to capture users' actions and motions. Moreover, VR requires additional feedback regarding the user's viewpoint. A user's viewpoint determines which tiled videos to deliver to render the FoV of the user. The viewpoint can be captured by motion tracking modules on a VR device. Additionally, motion emulation can be used to simulate a user's viewpoint movement on VR devices. VR devices can request the content proactively based on the emulation results to avoid performance degradation, such as rebuffering \cite{oculus}.  In addition, for interactive applications such as XR gaming, the sensors connected to XR devices, such as inertial measurement units (IMUs), haptic gloves, etc., gather inputs from the users. Depending on the inputs, the XR devices can either process the inputs locally or upload the inputs to content servers for computing and updating.

\subsubsection{Requirements}\label{sec.XR_req}

In general, XR has stringent latency requirements for accurate and smooth content playback based on user motions. In terms of VR, motion-to-photon (MTP) delay is the most important delay metric, which measures the time difference between the user's viewpoint movement and corresponding reflections at the output of the VR headset. If the MTP delay is larger than 20 ms, VR users may feel spatially disoriented and dizzy, referred to as VR sickness  \cite{oculus}. Current VR industries target lower MTP delay (below 15 ms) for ideal user experience \cite{Mangiante}. In addition, for VR applications requiring extensive interactions, the requirement of response time for rendering the interactions into VR content can be longer than the MTP delay requirement. For example, in VR gaming, a latency of up to 50 ms for responding to player actions can be noticeable yet currently acceptable \cite{Wuyang}. In terms of AR, the content is mainly captured by local devices. The MTP delay in AR can be minimized by playing the raw content captured by AR devices before the content is processed. However, users' immersive experiences can be adversely affected by delayed processing for rendering the user's motions into AR content. The delay requirements for reproducing user interactions in AR content are 75 ms for online gaming and 250 ms for telemetry based on the sensitivity of the human vestibular system \cite{mohan2020pruning}.

{Furthermore, in order to achieve low content delivery latency, an ultra-high data transmission rate is required for delivering XR content. Specifically, users view VR videos on headsets placed a few centimeters from their faces. Therefore, high-resolution videos are required for VR applications to improve user experience.
Although tile-based content transmission can reduce the data size in VR content delivery, data rate requirements can still be 2.35 gigabits per second (Gbps) or above for VR video delivery, which is more than 100 times higher than the data rate for current high-definition video streaming \cite{Mangiante}. For interactive XR applications, such as VR gaming and AR, extensive video processing is required. The computing capability of both network servers and user devices dominates the performance of interactive XR applications, and limited computing capability in the network can be another bottleneck for XR content delivery \cite{Elbamby}. }

\subsection{Haptic Communication}

In this subsection, we first provide the concepts of haptics and haptic communication. Then, we detail the implementation procedures and service requirements of haptic communication in the 6G era. 

\subsubsection{Concept}

The term haptics initially referred to interactions between humans and objects in the physical world that involve the sense of touch, e.g., swiping a phone screen~\cite{steinbach2012haptic}. The development of tele-operation technologies over the past few decades have expanded the definition of haptics to all forms of interactions involving the sense of touch, including interactions between humans and virtual objects in the virtual world or the tele-operated machines in the physical world~\cite{tan2020methodology,o2008haptic}. The information conveying the sense of touch in such interactions is referred to as haptic information. The sense of touch relates to different types of mechanoreceptors in human skin and muscles, and the haptic information can be broadly classified into tactile and kinesthetic information~\cite{abiri2019multi}. Specifically, tactile information is related to the sense of surface texture, friction, and temperature felt by the human skin when in contact with objects, and kinesthetic information is related to the sense of position and motion of limbs along with the associated forces~\cite{srinivasan1997haptics,steinbach2012haptic}. A device that supports haptic interactions and the transmission of haptic information is referred to as haptic interface (HI) or haptic device. \bl{Existing HIs can be broadly categorized into graspable, wearable, and touchable HIs. Generally, graspable HIs are mainly used for capturing and displaying kinesthetic information; wearable HIs are mainly used for capturing and displaying tactile information; and touchable HIs can be used in both kinesthetic and tactile information capture and display~\cite{culbertson2018haptics}.} An HI is comprised of haptic sensors and haptic actuators responsible for capturing and displaying haptic information, respectively~\cite{antonakoglou2018toward}. An HI can capture and display a variety of haptic information, and the number of independent coordinates used by the HI to specify the haptic information is referred to as the degrees of freedom (DoF) of the HI~\cite{promwongsa2020comprehensive}.

Haptic communication refers to the process in which humans communicate and interact through the sense of touch over a communication network~\cite{steinbach2012haptic}. The communication network supporting haptic communication is named as \emph{Tactile Internet} in some existing works~\cite{ali2022tactile}.\footnote{Haptic communication and the Tactile Internet are related as a service and a medium as in the case of voice over IP (VoIP) services and the Internet~\cite{aijaz2016realizing}.} With the use of HIs and the transmission of haptic information over communication networks, users can interact with virtual objects in the virtual world or remotely operate machines in the physical world~\cite{steinbach2012haptic}. The transmission of haptic information can be unilateral, bilateral, or multilateral, depending on the number of users participating in the haptic communication. In the cases of one user manipulating a remote machine or two users interacting with each other, the haptic communication is unilateral (i.e., an HI either sends or receives haptic information) or bilateral (i.e., an HI both sends and receives haptic information). In other cases, haptic information can be transmitted multilaterally, e.g., in cooperative tele-operations involving multiple users. This is because the behavior of each user may have an effect on other users, resulting in interconnections and couplings in the exchanges of haptic information~\cite{shahbazi2018systematic,feth2009shared}. Since haptic communication centers on humans, some studies examine the \emph{human-in-the-loop} nature of haptic communication and predict a paradigm shift from content delivery to skillset delivery, as a result of the emergence of haptic communication~\cite{simsek20165g,ali2022tactile}.

\subsubsection{Basic Implementation Procedure}

\begin{figure*}[htbp]
		\centering
	  	\includegraphics[width=1\textwidth]{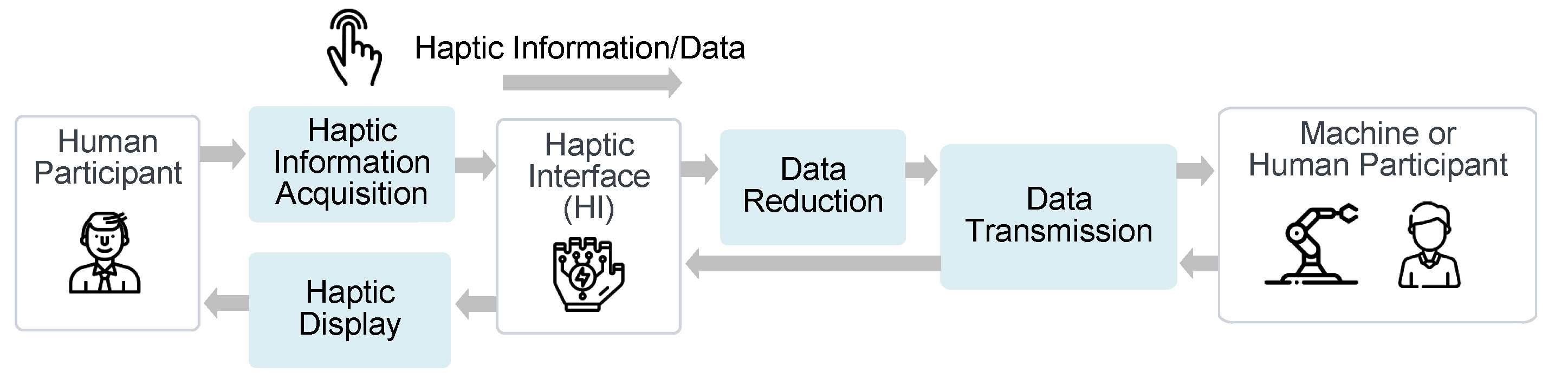}
	  	\caption{Implementation procedure of bilateral haptic communication.}\label{fig:hap_1}
\end{figure*}
\begin{figure*}[htbp]
		\centering
	  	\includegraphics[width=1\textwidth]{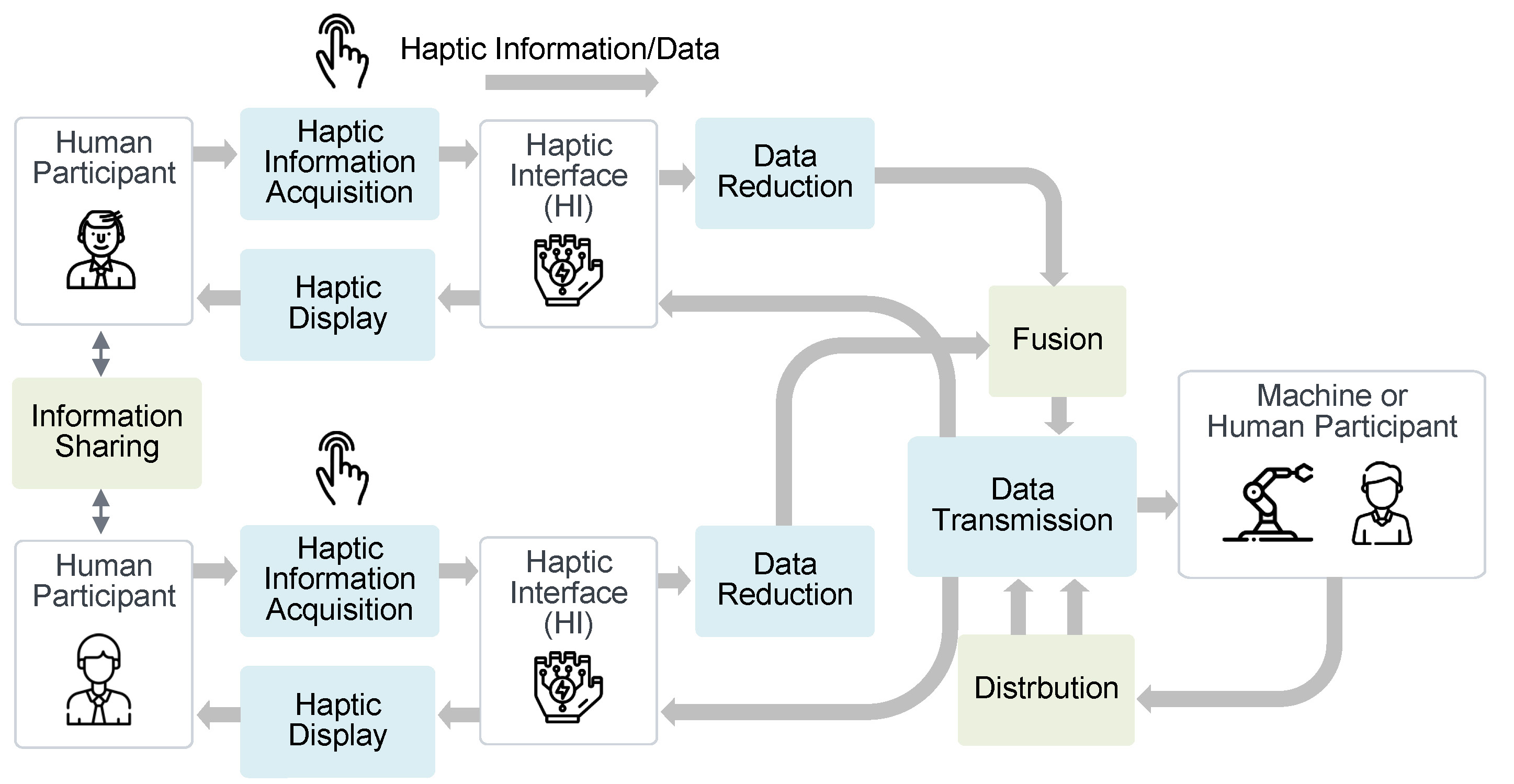}
	  	\caption{Implementation procedure of multilateral haptic communication.}\label{fig:hap_2}
\end{figure*}

The implementation procedures of haptic communication depend on how the haptic information is transmitted. For bilateral haptic communication, the implementation procedures mainly consist of four steps: haptic information acquisition, data reduction, data transmission, and haptic display, as shown in Fig.~\ref{fig:hap_1}.\footnote{In unilateral haptic communication, either the step of haptic information acquisition or the step of haptic display is skipped depending on whether an HI is sending or receiving haptic information.} In the first step, haptic information, including tactile and kinesthetic information, can be acquired by haptic sensors in HIs. In terms of tactile information, force sensors, thermistors, and laser scanners are mainly used in the measurement or evaluation of friction and hardness, warmth, and macroscopic roughness, respectively~\cite{liu2017challenges,lederman2009haptic,okamoto2012psychophysical,fishel2012bayesian}. 
Haptic sensors such as IMUs are responsible for the acquisition of kinesthetic information, e.g., tracking the position, velocity, and angular velocity of sensors positioned at different parts of a human~\cite{steinbach2018haptic}. The haptic sensors of interest can be dynamically selected, and only the haptic information captured by the selected haptic sensors needs to be collected for efficient haptic information acquisition~\cite{van2017challenges}. Due to the potentially high DoF of an HI, data reduction is adopted in the second step to reduce the amount of haptic data without degrading the users’ immersive experience too much. Specifically, waveform-based representation and feature extraction algorithms can be used in the compression of tactile information, and perceptual coding techniques based on perceptual masking phenomenon can be applied for compressing kinesthetic information~\cite{jayasankar2021survey,steinbach2010haptic}. In addition, predictive methods (also called predictive coding techniques) can be leveraged to reduce the amount of transmitted haptic data by inferring upcoming haptic information~\cite{steinbach2018haptic}. Haptic data reduction can be carried out at either HIs or network servers~\cite{fitzek2021tactile,steinbach2012haptic}. Existing methods of haptic data reduction are detailed in Section~\ref{hap}. In the third step, the haptic data can be transmitted over a communication network, resulting in a haptic data stream between two HIs. The haptic data stream can consist of multiple haptic data substreams, each of which corresponds to a type of haptic information. Data traffic patterns and QoS requirements can vary across different haptic data substreams due to the differences in the sensitivity of human perception, such as reaction time and the range of perception~\cite{fitzek2021tactile}. The respective QoS requirements of haptic data substreams should be satisfied, and the haptic data substreams should be synchronized in transmission. Moreover, a haptic data stream should be synchronized with audiovisual data streams in the case of immersive communications involving multiple modalities~\cite{cizmeci2017multiplexing}. In the last step, i.e., haptic display, haptic actuators in an HI stimulate human mechanoreceptors to create realistic haptic sensations when the HI receives haptic data~\cite{wang2019multimodal}. In general, haptic display includes tactile display, e.g., adjusting the temperature, and kinesthetic display, e.g., creating motion and changing muscle tension~\cite{ozioko2020wearable,steinbach2018haptic,pacchierotti2017wearable}. In the case when haptic data transmission is unreliable or delayed, predictive methods can be leveraged at the receiver side to estimate the haptic data not received timely for smooth haptic display. 

In the case of multilateral haptic communication, three additional steps take place besides the aforementioned four steps, especially for cooperative tele-operation applications~\cite{feth2009shared}. The implementation procedures of multilateral haptic communication are shown in Fig.~\ref{fig:hap_2}, and the three additional steps are highlighted with green rectangles. First, even if there is no direct haptic interaction between two users, they can still share haptic information~\cite{takagi2017physically}. \bl{For example, the information on tensile strength, texture, and depth of the tissue can be shared among surgeons to facilitate their collaboration in telesurgery.} The data format and content of the transmitted haptic information in such haptic information sharing may differ from those of the transmitted haptic information in direct haptic interactions~\cite{shahbazi2018systematic}. Second, it is necessary to properly fuse the haptic information from multiple users, e.g., the weighted sum, when their behaviors affect 
other users~\cite{thanh2012collaborative,fujimoto2008influences}. Third, when one user's behavior affects multiple users at the same time, distributing haptic information to multiple users according to their different behaviors is required to achieve precise haptic display for individual users, e.g., different reaction forces are applied to tele-operators~\cite{chen2016analysis}.

\subsubsection{Requirements}\label{hap:req}

The data transmission rate requirement of haptic communication is determined by the packet rate and size of haptic data.
The packet rate is the number of packets transmitted by an HI per second, which depends on the information update rate.
For the smoothness and fidelity of haptic perception, haptic information typically needs to be updated at a rate above 1,000 times per second \cite{1389970}. If each update of haptic information is packetized and transmitted, the corresponding packet rate of haptic data is above 1,000 packets per second \cite{xu2015haptic}. The packet size of haptic data largely depends on the DoF of the haptic data \cite{holland2019ieee}. For kinesthetic data,  controlling one movable component (e.g., a joint) on a tele-operator (e.g., a robotic arm) needs six coordinates to be specified to achieve 6 DoF, with three coordinates specifying the transitional motion in the 3D space and the other three specifying the rotational motion including roll, pitch and yaw, respectively \cite{promwongsa2020comprehensive}. Since a human hand consists of multiple movable components (e.g., finger joints and wrist joints), its kinesthetic data can be described by a 24-DoF model \cite{cobos2008efficient}. In addition, for reproducing tactile information with high fidelity, a dense array of haptic sensors/actuators needs to be deployed on a user \cite{hoggan2007mobile}. For example, for reproducing vibrotactile data, four actuators are deployed around one fingertip \cite{baik2020multi}. As a result, tactile data can involve even higher DoF than kinesthetic data \cite{holland2019ieee}. The packet size of 1-DoF, 10-DoF and 100-DoF haptic data is about 8, 80 and 800 bytes, respectively, and the specific data transmission rate requirement can be derived accordingly \cite{holland2019ieee}. 

The delay tolerance of haptic communication can be as low as 1 ms since the packet rate of haptic data can be above 1,000 packets per second \cite{fettweis2014tactile}. 
In practice, the delay requirement of haptic communication is determined by factors including the perceptual sensitivity of receivers, the dynamics of haptic interaction, and specific operation or interaction. First, higher perceptual sensitivity for haptic information generally indicates the need for a higher packet rate and thus a stricter delay requirement \cite{chaudhuri2018kinesthetic}. For example, while touring a virtual museum of natural history, archaeologists can have a stricter delay requirement than the majority of visitors due to their higher perception sensitivities of artifacts and specimens. Second, similarly, higher dynamics of haptic interaction generally call for a higher packet rate and a lower delay. Specifically, the delay requirement when such dynamics is high (e.g., in tele-soccer), medium (e.g., in telerehabilitation) and low (e.g., in tele-maintenance) is 1-10 ms, 10-100 ms and 100-1,000 ms, respectively \cite{holland2019ieee}. Moreover, for the same use case, the delay requirement can vary with the dynamics of the interaction. For example, in tele-training, the delay requirement when the trainee is being assessed and corrected by the trainer is 1-10 ms; the delay requirement when the trainee is observing the illustration of the trainer is 1-100 ms \cite{holland2019ieee}. 
Third, a delay below 2 ms is required for remote machine manipulation, while a delay below 50 ms is required for remote machine monitoring \cite{aijaz2018tactile}. 

The reliability of haptic communication can be evaluated in terms of bit error rate, packet loss rate, delay-bound violation probability, or prediction error when haptic data prediction is adopted \cite{promwongsa2020comprehensive}. The requirement for the reliability depends on factors such as the specific communication scenario and whether or not haptic data reduction is used. 
First, in terms of delay-bound violation probability, the reliability of haptic communication in immersive gaming is required to be above 99.9\% \cite{holland2019ieee}. In contrast, when critical operation tasks are performed based on haptic information, higher reliability of haptic communication is required. For example, the reliability of above 99.999\% is required for haptic communication in telesurgery and remote machine manipulation, \cite{gupta2019tactile, aijaz2018tactile}. Second, when haptic data reduction is adopted, the same packet loss or bit error rate can cause more degradation in the haptic information \cite{steinbach2010haptic}. As a result, the use of haptic data reduction can result in a stricter requirement for the reliability of haptic communication. For example, the reliability above 99.999\% is required in immersive gaming when haptic data reduction is adopted \cite{holland2019ieee}.

\subsection{Holography and Holographic Communication}\label{sss:Holo}

In this subsection, we introduce holography and holographic communication, beginning from presenting the concept and different types of holography, followed by the basic implementation procedure of holographic communication, and ending with the data transmission rate and delay requirements.  

\subsubsection{Concept}
As the name suggests, holographic communication depends on holography technology, which has made significant progress in the past decade. There are different stages in the development of holography technology. \textit{Optical holography} generates holograms via recording and recreating optical wavefront, and the corresponding holograms are recorded interference patterns (e.g., on photographic emulsions) of an ``object wave'' and a ``reference wave''. When the recorded interference pattern is illuminated by the reference wave, a 3D light field  can be recreated using diffraction. The original idea of hologram was developed in 1940s, and real breakthrough was made in 1960s thanks to the development of laser \cite{gabor1972holography}. Later, with advances in electronic devices, \textit{digital holography} emerged, which uses image sensors to capture interference patterns. In digital holography, recording is done optically, while a 3D image is reproduced via numerical calculation of light wave diffraction using methods such as Fourier transform \cite{10.1093/jmicro/dfy007}. The latest development of holography is \textit{computer-generated holography}, in which both the interference pattern and the 3D image in display are generated digitally using a computer~\cite{10.1145/3378444}. With computer-generated holography, the object to be displayed does not have to be physically present, which yields great flexibility at the cost of high computational complexity~\cite{10.3389/fphot.2022.854391}. Despite of the advance in recent years, generating dynamic 3D holograms in real time is challenging. As a result, alternative approaches to displaying 3D images emerge, which are sometimes referred to as ``false holography''. Such approaches use glass panes or other ``tricks'' to create illusions of 3D images~\cite{10.1145/1275808.1276427, NewsArticle13Hologram}. Among the false holography techniques, \textit{volumetric display} has attracted significant interest in the field of computer-aided design and medical imaging~\cite{1492264}. Volumetric display, an umbrella term for many different techniques, renders volume-filling 3D images via the generation, absorption, and scattering of illumination in a confined space, e.g., a cube or cone~\cite{yang2016see}. The study of volumetric display is active with exciting experiments~\cite{smalley2018photophoretic}, and commercial products are also available~\cite{gibney2019star}. Other approaches to imitate 3D display include the use of multiple projectors and a human-size retroreflective cylinder~\cite{10.1145/3173574.3174096}.
For example, a circular multi-projector array can be implemented for a light field cylindrical display to differentiate perceived images from different viewing angles.

Based on either true holography or ``false holography'', holographic communication is about transferring data representing dynamic 3D images of physical objects over a network and displaying the objects in 3D at the receiver.\footnote{Note that the term ``holographic communication" is also used in the literature of massive MIMO and IRS but with a different and unrelated meaning~\cite{9475156}.} Integrating 3D data capturing, processing, transmission, and rendering, holographic communication is expected to enable exciting new services in 6G~\cite{strinati20196g, clemm2020toward}. At the moment, there is no consensus on the scope of holographic communication in the literature, and some researchers consider the transferring and rendering of 3D data in AR/VR as a type of holographic communication~\cite{Ericsson2022}. In this review, holographic communication refers to data transfer for autostereoscopic 3D display, i.e., 3D images that can be viewed by naked eye without the aid of eyewear or headsets and, ideally, are different when viewed from different positions, angles, or tilts. The 3D display at the receiver can be rendered via real holography, false holography such as volumetric display, or other techniques as long as the objective of autostereoscopic 3D display is achieved. Similar to existing multimedia communications, the content of holographic communication can be either generated in real time or recorded, and the communication mode can be unicast, multicast, or broadcast.

\begin{figure*}[t]
		\centering
	  	\includegraphics[width=0.7\textwidth]{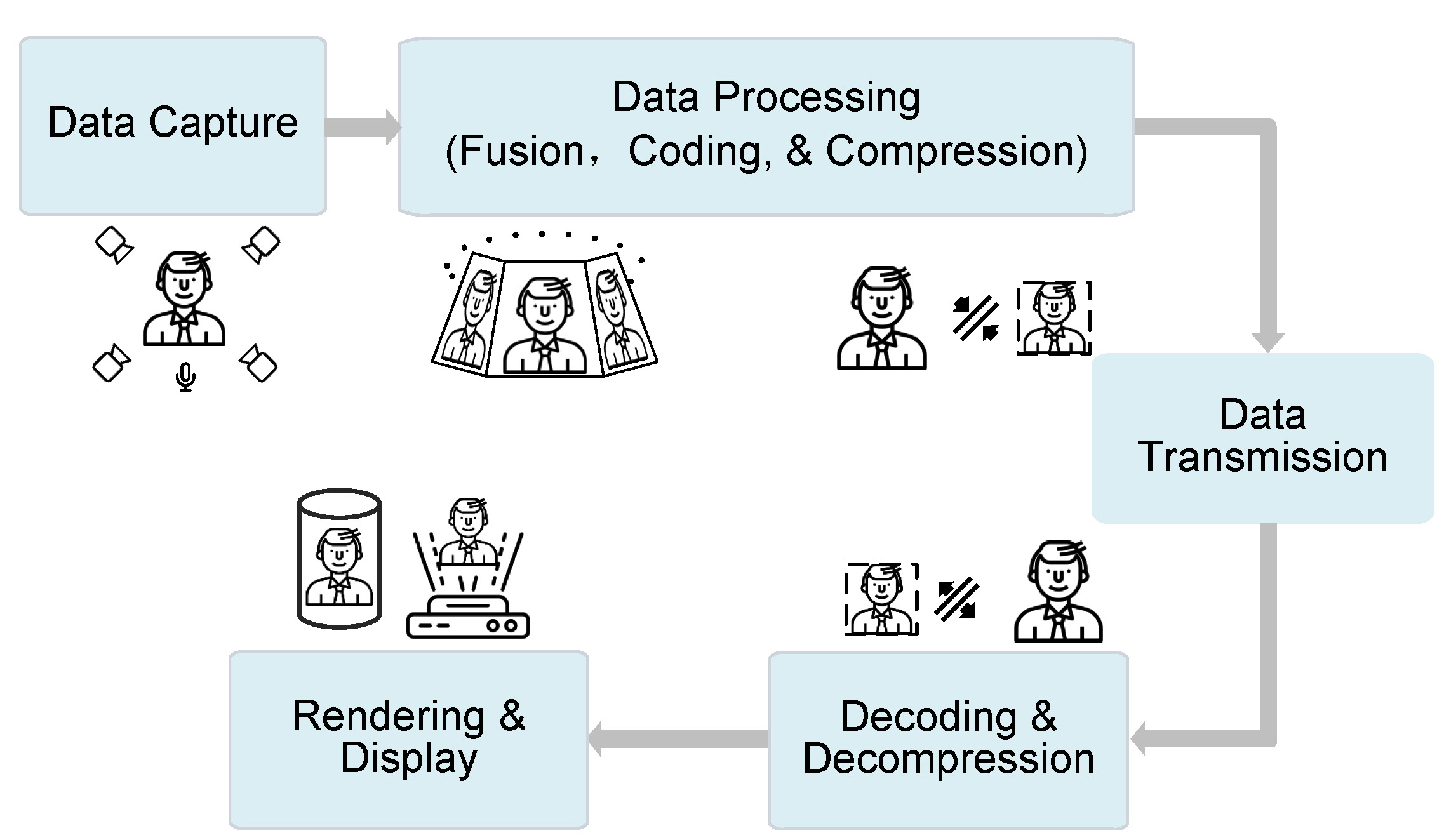}
	  	\caption{Implementation procedure of holographic communication.}\label{fig:hol_1}
\end{figure*}

\subsubsection{Basic Implementation Procedure}\label{sss:HoloProc}

Although various approaches for holographic communication differ in the implementation procedure, the general process includes the steps of data capture,  processing, transmission, and rendering, which is illustrated in Fig.~\ref{fig:hol_1}. 

Except for computer-generated holography, a capture system is required to record 3D images of a physical object. An ideal capture system for holographic communication would capture the light field, i.e., all the information of each light ray, in the target scene~\cite{apostolopoulos2012road}. In practice, capture is conducted with visual sensors such as a camera array \cite{10.1145/3305367.3327974} or light detection and ranging (LIDAR) sensors \cite{fratz2021digital}. The depth information of the object of interest is either directly captured (e.g., in the case of a capture system with LIDAR sensors) or computed in the subsequent data processing step (e.g., in the case of a capture system with a camera array). The performance of the visual capture system depends on factors such as the number of sensors and the camera sampling rate~\cite{apostolopoulos2012road}.  

In the data processing step, the depth information of target objects in the scene is computed (if not directly captured), and the output from capture sensors is fused to form a composite 3D representation of the captured scene \cite{Javidi05}. For example, in digital holography, a computer can process 2D images taken from different angles and tilts by a camera array to form a single 3D representation of the captured scene~\cite{Ericsson2022}. The fusion of images may help achieve visualization enhancement in the rendered 3D images such as improvement in the resolution and contrast~\cite{Javidi05}, and it can be conducted either solely at the transmitter side or with the help of an edge server.  In addition, the data processing step is responsible for the compression of the fused data to speed up the transmission and reconstruction, and reduce the required data transmission rate and storage in holographic communication~\cite{KURBATOVA2015328,cheremkhin2019wavelet}. The compressed data for the 3D representation is then encoded and transmitted over a network.

At the receiver side, the received data is decoded using one or multiple chosen codecs and decompressed. The captured scene is then reconstructed, possibly with the help of an edge server, and rendered on a display device. An ideal display device for holographic communication would regenerate the light field in the captured scene to create an illusion that the user is placed in the scene. In practice, creating such an illustration is difficult as it requires each point (e.g., each pixel) of the display device to emanate different light rays in different directions. However, given the limitations of human perception, the feeling of visual immersion can be created by using equipment such as a cylindrical light field display \cite{10.1145/3173574.3174096}, a persistence of vision (PoV) display \cite{gately2011three}, or a static volumetric display device \cite{kumagai2021colour}. Such devices render 3D images by using a large curved display to fill the user's FoV, exploiting the phenomena of a lingering afterimage on the retina, and dynamic turning on/off of voxels in a confined 3D space, among other methods for creating illusions of 3D images.      

It is worth noting that holographic communication may also involve audio data capture, processing, and rendering. In such a case, capturing the sound field in the target scene and ensuring audio and video synchronization are important for users to enjoy an immersive holographic communication experience~\cite{apostolopoulos2012road}.

\subsubsection{Requirements}

Holograms mainly come in two types, namely volumetric-based holograms and image-based holograms. 
The transmission of the two types of holograms requires different data rates, ranging from hundreds of Mbps up to Tbps \cite{clemm2020toward}.
For volumetric-based holograms, a physical object is represented as a set of 3D pixels or voxels, such as a point cloud.
Transmitting a point cloud targeting an object requires a data rate on the level of hundreds of Mbps to several Gbps, depending on the resolution of the 3D content \cite{fg2020representative}.
For example, to fully represent a human, the point cloud in each frame typically consists of $10^5$ to $10^6$ points, while each point needs 15 bytes of data to represent the color and 3D coordinate of the point. 
In the case of 30 frames per second, the data rate requirement is between 300 Mbps and 3 Gbps \cite{Ericsson2022, selinis2020internet}.
For image-based holograms, such as light-field video (LFV), an object is presented by an array of images captured at different angles, tilts, and/or positions.
An LFV-based hologram can be more precise as compared with a volumetric-based hologram, especially in high resolution when a large number of images from different tilts, angles, and positions are used per frame \cite{jiang2021road}. 
For example, if the 3D representation of an object requires a separate image every 0.3$^{\circ}$, a hologram with an FoV angle range of 30$^{\circ}$ and a tilt range of 10$^{\circ}$ needs 3,300 separate 2D images.
In order to transmit an LFV-based hologram for a human-sized object, the required data rate should be between 100 Gbps and 2 Tbps \cite{clemm2020toward}.

To support real-time holographic communication, the overall delay, including data capturing, processing, transmission, and rendering delay, should be less than 100 ms \cite{he2023three}. 
In addition to low delay, synchronization is important to holographic communication.
Generally, the hologram of objects or humans may be sampled by multiple sensors from different angles and different distances.
In this case, data from different sensors should be synchronized in transmission \cite{strinati20196g}.
Taking holographic teleconference as an example, as multiple participants can join the teleconference from different locations, multi-source synchronization is necessary for them to have good quality of experience (QoE) in holographic communication.
Otherwise, a part of the rendered hologram can be slightly ahead or behind relative to the rest of hologram for some users, resulting in poor QoE \cite{lesniak2018dynamic}.
Moreover, holographic communication can involve multi-sensory information, e.g., the haptic, audio, and video information \cite{taleb2021extremely}. 
In this case, the synchronization of different sensory information in transmission is also important for users to see the hologram, hear the voice, as well as receive touch-sensory feedback from others without a degradation of the immersive experience due to out-of-sync issues.
For holographic communication involving the transmission of audiovisual and haptic data, the tolerable difference in the delay of different types of data should be lower than 80 ms for satisfactory QoE \cite{montagud2018mediasync}.

\begin{table*}[hbt!]
    \caption{Requirements of use cases in immersive communications}\label{Table2}
    \renewcommand{\arraystretch}{1.5}%
    \begin{tabular}{c|p{7cm}|p{7cm}}
        \hline\hline
            ~ & \textbf{Use Cases} & \textbf{Requirements} \\ 
        \hline\hline
          \multirow{6}*{\textbf{XR}} & \multirow{2}*{360\textdegree~ video playback} & $<$ 20 ms MTP delay \cite{oculus}, 2.35 Gbps data rate \cite{Mangiante}\\ \cline{2-3}
          ~& \multirow{1}*{Interactive applications (e.g., VR gaming)} & $<$ 50 ms response time \cite{Wuyang}    \\ \cline{2-3}
            ~ & Collaborative virtual applications (e.g., teleconference) & $<$ 150 ms virtual feedback \cite{jay2007modeling}, 12.5 Tbps/km$^2$ upload capacity\\ \cline{2-3}
            ~ & \bl{AR smart healthcare} & \bl{ $<$ 5 ms delay, 10 Gbps data rate \cite{xu2022full}}\\ \cline{1-3}
          \multirow{3}*{\textbf{Haptics}}  &
            \multirow{1}*{Telesurgery }& $>$ 99.999\% reliability, $<$ 1 ms delay \cite{gupta2019tactile}   \\ \cline{2-3}
            ~ & \multirow{1}*{Remote machine manipulation} & $>$ 99.999\% reliability, $<$ 2 ms delay  \cite{aijaz2018tactile}\\ \cline{2-3}
            ~ & \multirow{1}*{Haptic interaction-based rehabilitation}  & $>$ 99.999\% reliability, $<$ 50 ms delay \cite{holland2019ieee}  \\ \cline{1-3}
          \multirow{4}*{\textbf{Holography}} &  Volumetric-based hologram (e.g., point cloud) & $>$ 300 Mbps data rate \cite{Ericsson2022, selinis2020internet}  \\ \cline{2-3}
            ~ & Image-based hologram (e.g., LFV) & $>$ 100 Gbps data rate \cite{clemm2020toward} \\ \cline{2-3}
            ~ & Real-time holographic teleconference & $<$ 100 ms delay \cite{he2023three} \\
        \hline\hline
    \end{tabular}  
\end{table*}

\section{Immersive Communications: Challenges and Solutions}\label{s:Solus}

\begin{figure*}[htbp]
		\centering
	  	\includegraphics[width=0.6\textwidth]{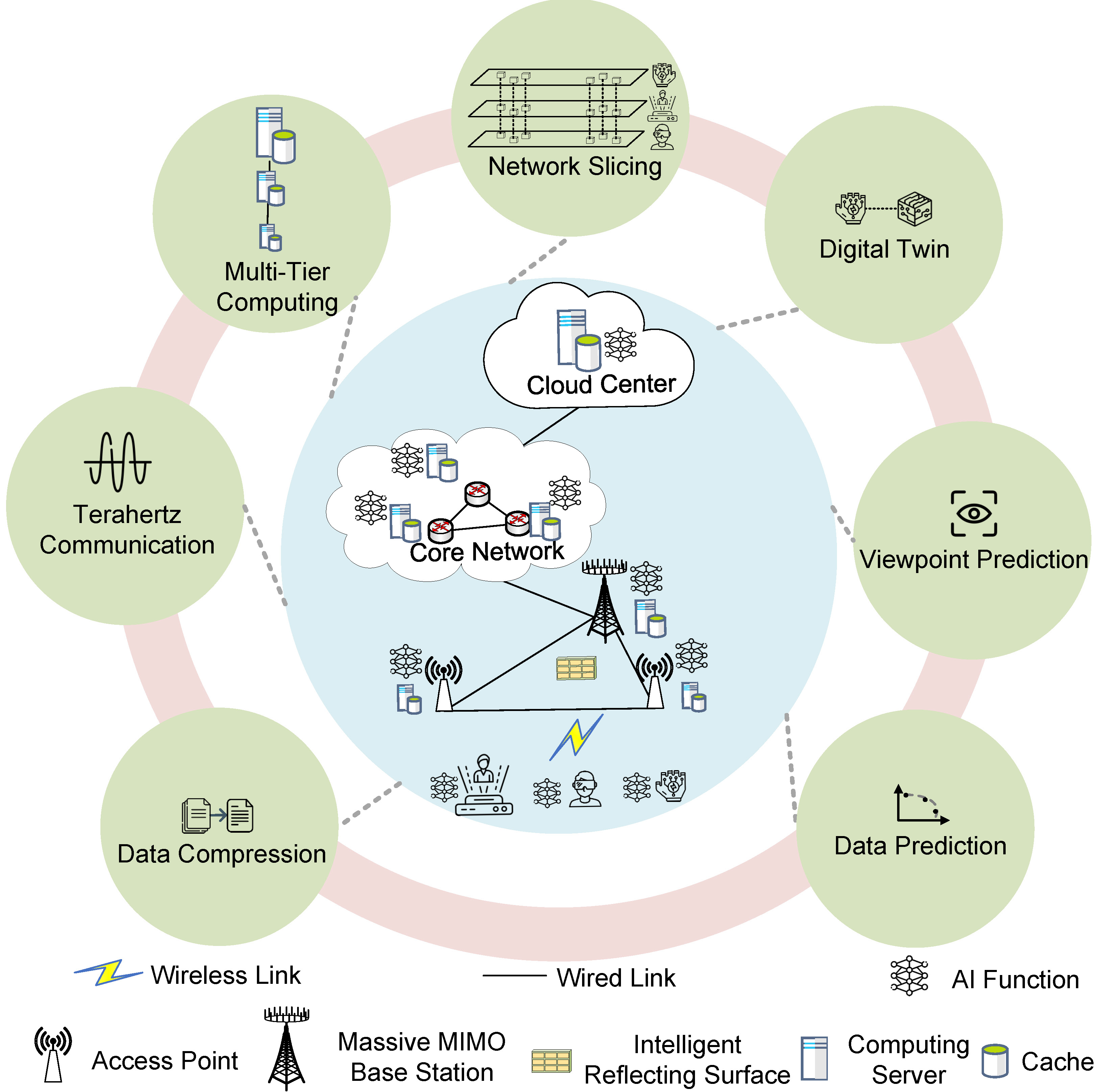}	  	
	  	\caption{Potential solutions to immersive communications.}\label{fig:2}
\end{figure*}

After introducing the concepts, implementation procedures, and requirements of immersive communications, we now discuss challenges in XR, haptic communication, and holographic communication, as well as the state-of-the-art solutions, with the most important ones summarized in Fig.~\ref{fig:2}. Note that our review here focuses on the challenges and solutions related to the communication, computing, and networking aspects of immersive communications.

\subsection{Extended Reality}

{The main challenge of XR is delivering the required content to users on time, given the limited transmission resources and computing capability in a network. A variety of network functions and resources contribute to the performance of content delivery. Systematic solutions involving data processing, rendering, transmission, etc., have been developed to address these challenges. We summarize the solutions for implementing XR in three aspects: content selection, transmission improvement, and computing optimization.}

\subsubsection{Content Selection}
{The fundamental step in supporting XR applications is to identify which content needs to be processed and transmitted. This step focuses on minimizing the overall data size of the content to deliver at the cost of tolerable performance degradation, thus reducing the delivery time.}

In VR services, proactive content delivery is commonly used to meet MTP delay requirements. Thus, in tile-based content transmission, the primary research challenge is how to predict user viewpoints accurately so as to determine which tiled videos to deliver to users. The prediction of user viewpoints can be achieved by sequential learning and data analysis methods based on the user's viewpoint trajectory, such as linear regression \cite{10.1145/3210240.3210323,Nasrabadi}, and long short-term memory (LSTM) \cite{10.1145/3229625.3229629}. A lightweight viewpoint prediction function can be deployed at the VR headset for local viewpoint prediction. Alternatively, the viewpoint trajectory can be updated to a network server (e.g., edge server), in which a more advanced machine learning model can be applied for accurate prediction \cite{9069299}. If the viewpoints are predicted by the network server, the prediction can be conducted based on not only current viewpoint trajectories for a group of users \cite{sun2020flocking} but also the historical viewpoint trajectory data to further improve the prediction accuracy \cite{Xu_2018_CVPR,feng2019viewport}.
Although viewpoint prediction enables proactive tile-based content delivery, perfect prediction cannot be achieved due to the dynamics of user viewpoint movement. Even if viewpoints are known in advance, dynamic network environments such as data traffic load and processing time require adaptive resource management to ensure playback performance. With stochastic decision-making methods, such as reinforcement learning, it is possible to identify the dynamics of user viewpoint movement and determine which tiled videos to deliver to the corresponding VR device \cite{9614984}. In addition, the portion of tiled videos with different video qualities transmitted in a given time interval can be adjusted according to the viewpoint movement of a user. Increasing the portion of low-quality videos can improve the robustness against viewpoint prediction errors, while increasing the portion of high-quality videos can improve the QoE of the user. The optimal tradeoff between the robustness and the QoE is evaluated for VR video delivery in \cite{9939105}. 

{AR devices capture raw content, i.e., video frames, which can be offloaded to network servers for prompt content processing. Once the raw content is offloaded, the server detects and processes the objects within video frames captured by users' cameras, then returns the processed content to the AR devices. Though it is easier to satisfy the MTP delay requirement in AR than VR, enabling accurate and rapid content processing (e.g., object detection) by network servers requires sufficient bandwidth to provide low-latency two-way transmission for satisfactory QoE. To balance transmission bandwidth usage for computing offloading and content processing performance, current solutions mainly focus on using machine learning techniques to adjust the number of frames offloaded by an AR device per unit time, based on the network environment and AR device movement. Specifically, offloading more video frames to a network server can improve object detection accuracy, especially when the AR device moves quickly and generates new content frequently. However, the bandwidth usage increases accordingly due to a large number of frames to offload \cite{liu2018edge}. Taking AR device mobility and network dynamics into account, adaptive frame rate adjustment is investigated in \cite{9246260}. A deep reinforcement learning approach is used to study how mobility dynamics affect AR service performance and to determine the optimal uploading frame rate for maximal object detection accuracy and playback fluency. }

{XR content is expected to be further enriched in the era of 6G. Digital twins can incorporate AI to collect environmental information, characterize physical objects, and construct digital models of the physical objects accordingly. Digital models from digital twins can be used for XR applications as a new type of XR content that can be accessed by XR devices \cite{zhang2022artificial}. For example, in an industrial Internet-of-Things scenario, designers and workers can use XR devices to interact with the digital models of machines and products in a simulated virtual environment. In addition, XR devices can collect the interactions from designers and workers. Based on the interactions, digital twins can adaptively configure their settings, such as data collection frequency \cite{aheleroff2021digital}. The combination of XR and digital twins can support emerging applications such as metaverse. However, synchronizing among the physical world, digital twins, and XR content requires considerable network resources. Game theoretic methods are adopted in \cite{9865226} to adjust the synchronization rate between the physical world and digital twins based on the demand of virtual service providers that provide content to XR devices. A network slicing-based solution is proposed for providing metaverse services \cite{liu2022slicing4meta}, which allocates multi-dimensional resources for content synchronization to improve the fidelity of digital twins and the QoE of XR users.}

\subsubsection{Transmission Improvement}
{As discussed in Subsection \ref{sec.XR_req}, the main bottleneck for VR video delivery is a limited data rate. Therefore, a straightforward solution to overcome the bottleneck is to increase the data rate with advanced communication techniques. As a key technology in 5G, millimeter wave (mmWave) communications can facilitate VR content delivery due to their high data rate and ultra-low propagation latency \cite{abari2016cutting}. In 6G, the transmission rate can be further improved by the physical layer technologies of terahertz (THz) transmission and intelligent reflecting surface (IRS), which can be applied in VR video\cite{9149411, 9120235}.  
However, communication links using ultra-high frequency bands, such as mmWave and THz, are prone to outage as they require line-of-sight (LoS) channels. Physical obstacles in the environment, including the user's body, may break the communication links and severely degrade the communication quality. To address this issue, a sub-6 GHz frequency band can be used as a backup if the mmWave or THz bands does not provide satisfactory channel quality. However, dynamic frequency band switching can result in a time-varying data transmission rate, thereby degrading the content delivery performance. The work \cite{Yanwei} models communication link state transitions corresponding to switching different frequency bands (e.g., mmWave and sub-6 GHz bands) in VR content delivery as a Markov chain. Content processing policies are adjusted to compensate for transmission delays when channel state transitions occur. In addition to adapting to channel dynamics, the reliability of mmWave or THz communication links can be improved by establishing multiple communication links between a device and several edge servers for VR content delivery \cite{9798771, 9536410}.} \bl{In addition, on the link layer, IEEE 802.11 releases a new amendment
standard IEEE 802.11be – Extremely High Throughput (EHT), i.e., WiFi-7, to support high-throughput and low-latency video applications, including XR, through aggregating multiple transmission bands, exploiting MIMO enhancements, and enabling multi-AP coordination \cite{9152055}. }

On the network layer, a network virtualization-based solution is proposed for VR content delivery, in which network controllers can create private logic networks for VR applications to satisfy their service requirements and dynamically adapt the routing schemes according to the mode of content delivery (i.e., uni-cast or multi-cast) \cite{huawei_vr}. The transmission protocols are designed according to the features of VR content delivery. The transmission protocol based on quick UDP Internet connections (QUIC) is proposed in \cite{yen2019streaming} to prioritize important tiled videos, such as the videos in the center of the user's FoV or the videos to be played soon, in transmission over a QUIC connection, in order to minimize the ratio of missing tiles when playing VR videos.

\subsubsection{Computing Optimization}
{Supporting wireless XR requires networks to have sufficient computing capability for processing and rendering the content, especially for interactive applications such as VR gaming. Processing the content locally at the XR devices can be time-consuming and energy-inefficient due to their limited computing capability. Instead, the computing workload can be fully or partially offloaded to network servers, and multi-tier computing can be a potential solution to reduce computing time and bandwidth consumption when providing computing services to XR devices. Accordingly, computing strategies should base on the features of diverse network servers to improve resource utilization and service performance. }

In MEC, edge servers can provide additional computing capability for resource-limited devices to reduce content processing latency for mobile XR content delivery. Specifically, in VR, edge servers can project monoscopic videos to stereoscopic videos when content is transmitted from the content provider's cloud server to VR devices. Such MEC-assisted content delivery can reduce bandwidth consumption compared to delivering stereoscopic videos from the cloud server directly, and computing time can be reduced compared to projecting the videos at the local devices \cite{Mangiante}. In AR, devices can offload captured content to an edge server to minimize processing latency \cite{9363323}.  In addition, edge servers can cache the processed XR content to further reduce the content delivery and processing time \cite{sukhmani2018edge}. Joint computing, caching, and communication resource management for VR video delivery is investigated in \cite{Yaping, Dang}, which studies the tradeoffs between computing and caching resource allocation for minimizing content delivery delay, given stochastic content processing time and popularity. Deep reinforcement learning methods are adopted to allocate computing resources at an edge server for individual content delivery requests in \cite{Yanwei, 9411714}, aiming to minimize content delivery delay while adapting to dynamic network environments and user viewpoint movement. \bl{The work \cite{Siyi} further investigates trusted caching collaboration for multiple edge servers in supporting VR/AR content delivery. A distributed caching scheme is proposed to optimize the cache space and policy for edge servers while incentivizing edge servers to participate in edge caching through verification schemes in the blockchain.}

Nonetheless, the computing capability at edge servers may not always be sufficient for processing XR content. Compared to cloud servers, edge servers usually have limited storage resources for caching XR content. Targeting 6G, a multi-tier computing architecture provides a potential solution for further accelerating XR content delivery by coordinating computing and storage resources among cloud servers, fog servers (e.g., servers at the gateway), and edge servers across the network. By integrating computing resources across the entire network, content processing workloads can be optimally distributed among multiple servers, and storage capacity among servers can be utilized to satisfy offloaded computing demands. However, optimizing XR performance by multi-tier computing can be complicated when there are a multitude of computing offloading and caching options to choose. The computing and caching resource coordination between the cloud server and edge servers is studied in \cite{mehrabi2021multi} and \cite{8673791}. Based on the information of a static network environment, e.g., transmission rate and XR computing demand, mixed integer nonlinear programming is investigated. Considering dynamic network environments and user mobility, the work \cite{zhou2022digital} utilizes digital twins of end users to characterize network dynamics and statuses. The meta-learning method is adopted to jointly allocate computing and caching resources at servers on different tiers of a network for context-based applications, including XR, based on the captured network statuses from digital twins. \bl{The attention of users on the virtual objects in XR content is predicted in \cite{du2022attention} by an alternating least square method, and a computing resource allocation scheme is proposed to prioritize processing of the virtual objectives that attract more user attention.}

{In addition to jointly allocating computing and caching resources at network servers,  computing performance can be further enhanced by scheduling computing tasks at edge servers. Edge servers can provide location-based content to users, which can contribute to computing optimization for XR applications. Specifically, in AR, users at close locations may offload and require similar content, and therefore, raw content offloaded from the nearby users can be processed together for improving computing efficiency \cite{jia2018delay}. Furthermore, rendering pipelines can be optimized based on real-time communication and computing performance of network servers and local devices when part of the workloads for content rendering are offloaded. A collaborative rendering pipeline is investigated in \cite{xie2021q}, which dynamically arranges the execution order of sub-tasks in content rendering on both the edge server and XR devices, based on network characteristics, to facilitate parallel computing and improve content rendering efficiency.} \bl{In addition, joint computing and communication resource management for efficiently supporting multiple users in a virtual world is investigated in \cite{9300168}. Device-to-device links are enabled to allow each AR device to leverage the computing resources of nearby AR devices for lightweight pre-processing of the captured frames to further improve computing resource utilization in the network. }

\subsection{Haptic Communication}\label{hap}

The main challenge in haptic communication is to satisfy the stringent delay and reliability requirements in the delivery of haptic data, especially when the data packet rate is high. To tackle this challenge, solutions have been developed in three aspects, including haptic data reduction to reduce the packet size or the packet rate, advanced communication and networking techniques to reduce delay and improve  reliability, and haptic data prediction to compensate for  excessive delay and packet loss over communication networks.

\subsubsection{Haptic Data Reduction}\label{hap_data_red}

To improve the fidelity of haptic perception, the number of haptic sensors/actuators deployed on an HI has been increasing \cite{steinbach2018haptic}. For example, electronic skin (e-skin) can be attached to prosthetic limbs for sensing haptic information, or to human skin for virtual social interaction \cite{yu2019skin, dahiya2019skin}. To reproduce the function of human skin, sensors/actuators need to be densely deployed on e-skin, for example, 25 sensors/actuators per 1 $\rm{cm}^2$ \cite{liu2020electronic}. In addition, the required packet rate for haptic data can be higher than 1,000 packets per second \cite{ orlosky2017virtual}. As a result, with a large number of devices and a high packet rate, the required data transmission rate of haptic communication can be high. To tackle this challenge, one solution is haptic data reduction, which is to reduce the packet size or rate of haptic data.

For reducing the packet size of haptic data, floating-point compression in the time domain or quantization of haptic data in the frequency domain can be exploited. In floating-point compression, one degree of freedom in the haptic information (e.g., the direction of the transitional movement in an axis) can be represented by a 32-bit floating-point number, and only the bits different from those in the previous haptic data are transmitted \cite{you2008haptic}. Using time-frequency transformation algorithms such as discrete cosine transform, a sequence of haptic data packets in the time domain can be transformed into the data in the frequency domain, which are then quantized and transmitted \cite{4957156, zeng2020perception}. For reducing the packet rate of haptic data, the perceptual masking phenomenon is widely exploited, which suggests that a human cannot perceive the difference of haptic information below the just-noticeable difference (JND). According to the Weber’s law, the JND of haptic information is proportional to the currently perceived value of the information, and the proportion is referred to as the Weber fraction \cite{steinbach2018haptic}. In this regard, the perceptual haptic reduction method is to transmit an updated haptic data packet only when the difference is larger than a threshold (e.g., JND) \cite{steinbach2010haptic}. 
In addition, the perceptual masking phenomenon in both time and frequency domains can be jointly exploited to achieve a higher data reduction ratio and lower data deviation \cite{wei2022perception}. Moreover, by jointly evaluating the difference of the haptic information in terms of all the DoF among consecutive data packets, the perceptual haptic reduction can be further improved \cite{steinbach2012haptic}.

The use of haptic data reduction should adapt to the type of haptic data, the delay requirement and the reliability requirement for haptic communication. First, haptic data can exhibit different Weber fractions in the JND, e. g., $7\% \thicksim 15\%$ for force data and $13\% \thicksim 28\%$ for stiffness data, which results in different thresholds in perceptual haptic reduction \cite{chaudhuri2018kinesthetic}. Second, data reduction in the frequency domain results in high processing delay since it is based on a sequence of data packets in the time domain. It is suitable for use cases with high delay tolerance, such as the passive perception and exploration of remote/virtual objects \cite{sachs2018adaptive}. In contrast, data reduction in the time domain, implemented in real time, is suitable for use cases with low delay tolerance such as immersive gaming, which involves extensive interactions between the players \cite{holland2019ieee}. Third, haptic data reduction may not be suitable for use cases requiring high reliability. As discussed in Subsection ~\ref{hap:req}, with the use of haptic data reduction, the required reliability of haptic communication increases. In this regard, for use cases with a high-reliability requirement (e.g., $99.999\%$ for telesurgery), the reliability requirement can be difficult to satisfy if haptic data reduction is used.

\subsubsection{Communication and Networking Solutions}\label{haptic_sol}

To satisfy the ideal communication delay of below 1 ms for haptic communication, physical-layer delay of less than 0.1 ms is desired \cite{aijaz2016realizing}. For reducing queuing delay, haptic data may be allowed to preempt the data of other types in the downlink transmission \cite{ji2018ultra}. For uplink transmissions, non-orthogonal multiple access (NOMA) can improve spectrum efficiency and reduce channel access delay of haptic devices \cite{8611157}. In addition, a grant-based user scheduling mechanism can take 0.3-0.4 ms for exchanging the scheduling request and transmission grant \cite{ji2018ultra}. Besides such delay, the signaling overhead, resulting from network control or grant-based scheduling, reduces the efficiency of data transmission \cite{ding2021enabling}. Therefore, grant-free user scheduling has been exploited to avoid the time-consuming scheduling, which periodically pre-reserves transmission resources, and the same resources can be pre-reserved to multiple haptic devices for improving resource utilization \cite{ali2021urllc, gao2021mac}.
To reduce the delay due to packet retransmissions, interference in multiple access should be properly managed. In grant-free NOMA, the interference can be managed by device activity detection \cite{ye2019deep} and successive interference cancellation (SIC) \cite{8533378}. Rate-splitting multiple access (RSMA) encodes message streams intended for multiple devices into common streams and private streams based on available channel state information (CSI), and a device  jointly decodes the common streams and the private stream intended for it, which can achieve flexible interference management and high robustness to imperfect CSI \cite{9348672}.

For improving the communication reliability of haptic communication, several approaches have been adopted in the literature. First, considering the small size of a haptic data packet, short block-length channel codes with strong error correction capabilities, such as low-density parity-check (LDPC) codes and short polar codes, have been investigated for haptic communication \cite{yuan2022exploring, miloslavskaya2020design}. Second, spatial diversity can be exploited by massive multiple-input and multiple-output (MIMO), IRS, and multi-connectivity techniques \cite{tarneberg2017utilizing, tang2020mimo, anwar2021physical}. 
Third, time diversity can be exploited by retransmission schemes such as $K$-repetition, in which a haptic device can automatically transmit $K$ repetitions of a packet over consecutive slots, thereby avoiding the delay caused by waiting for a retransmission request from the receiver \cite{yang2021heterogeneous}. NOMA can improve retransmission efficiency where the transmit power of a device can be optimized to retransmit the required minimum redundant bits for satisfying the reliability requirement \cite{8761473,9254087}.

To guarantee low delay and high reliability for haptic communication, network slicing, which allows multiple isolated virtual networks to be constructed over a shared physical network infrastructure, has been exploited \cite{polachan2020dynamic}. 
The perceptual masking phenomenon of haptic information, as introduced in Subsection ~\ref{hap_data_red}, can be exploited to accurately capture the maximum tolerable delay of haptic communication requests, which facilitates resource reservation in the network slice for haptic communication \cite{ge20195g}. For multiple tele-operation slices, diverse stability control capabilities of tele-operators in the presence of delay should be considered for customized transmission resource reservation \cite{liu2018qoe}. Moreover, by exploiting AI-based learning methods, traffic patterns of haptic devices can be accurately captured, and efficient resource reservation can then be facilitated \cite{shen2020ai}.

\subsubsection{Haptic Data Prediction}
The delay requirement of haptic communication can impose a constraint on the distance between two users. For example, to satisfy a delay requirement of 10 ms, the distance between a transmitter HI and a receiver HI must be smaller than 3,000 km since the propagation speed is upper-bounded
by the speed of light. This can create an issue for applications such as VR gaming with haptic interactions of players across continents. In addition, it is impossible to eliminate the loss of data packets or the violation of delay requirement in haptic communication \cite{aijaz2018tactile}. 
To improve user experience considering the above facts, haptic data prediction can be exploited.

For haptic data prediction, model-based or model-free prediction algorithms can take historical haptic data and other correlated data as the input. In tele-operation, the force feedback from the tele-operator is predicted by evaluating the previous force feedback through an auto-regressive model \cite{4285132}. In the tele-operated needle insertion, the force/torque feedback from the patient is predicted by inputting the force/torque commands of the surgeon to the hidden Markov model (HMM) \cite{9045432}. Audiovisual data collected in the interaction with a surface material are input to a neural network-based semantic learning algorithm to predict the texture of the surface material \cite{wei2021haptic}.

Haptic data can be predicted either at the receiver side or at the transmitter side to compensate for an excessive delay or packet loss. The receiver can predict the haptic data from the transmitter when an excessive delay occurs \cite{maier2019towards}. For example, digital twin-based prediction can be used by the receiver for low-latency interactions \cite{el2018digital}. Alternatively, the transmitter can predict its future haptic data and transmit the predicted data to compensate for the transmission delay \cite{hou2019prediction}. In this case, the prediction of whether haptic interaction is about to occur can assist to determine whether the haptic data prediction and the subsequent transmission are necessary \cite{mondal2020enabling}.

Haptic data prediction algorithms, such as AI-based ones, can be computing-intensive. To this end, they can be implemented using computing resources in the network to satisfy the stringent delay requirements \cite{simsek20165g, sukhmani2018edge}. In a tele-operation scenario, each of the two interacting haptic devices is associated with one edge server which caches the haptic interaction data, trains and implements the LSTM network-based prediction algorithm, and delivers the predicted haptic data to its associated haptic device \cite{li2021edge}. Furthermore, with close proximity, auxiliary robots can be deployed around haptic devices to implement haptic data prediction and deliver the results to the devices using device-to-device (D2D) communications \cite{yu2022can}.

In addition to compensating for the delay or packet loss, haptic data prediction can be used to reduce the packet rate of haptic data \cite{antonakoglou2018toward}. Specifically, the haptic transmitter can implement the haptic data prediction and evaluate the prediction deviation, and only transmit the data when the prediction deviation is higher than the JND of the receiver. If the haptic data has not been transmitted, the receiver can predict it based on the prediction algorithm shared with the transmitter.

\subsection{Holographic Communication}
In holographic communication, users are able to view 3D holograms from different angles, tilts, and positions.
As a result, a hologram synthesized with information from more viewpoints can produce more detailed and continuous visual information for users, thereby creating a more realistic immersive experience \cite{liu2019network}. 
This however requires the transmission of a large amount of data.
The main challenge in holographic communication is its stringent data rate and delay requirements. 
In this subsection, we focus on potential solutions for tackling this challenge in the aspects of data processing, communication, and networking.

\subsubsection{Content Selection, Compression and Prediction}
A high data rate is essential for holographic communication, and the demand for data rate can vary from hundreds of Mbps to several Tbps depending on the type of transmitted data, e.g., volumetric-based or image-based holograms. 
One way to relax the data rate requirement is to reduce the data size, for example, by transmitting only the most essential parts of a hologram through viewpoint-based content selection in holographic communication \cite{clemm2020toward}.
Since some parts of the hologram may not be observed depending on the user's viewpoint and position, as well as the presence of obstacles, those parts may not need to be transmitted.
However, two issues remain even with the selective transmission.
First, for an immersive experience in holographic communication, 6 DoF (yaw, pitch, roll, up/down, left/right, forward/backward) need to be considered when a user views a hologram, which make content selection based on the user's viewpoint a complex problem.
In addition, without head-mount devices such as VR headsets, tracking the position and viewpoint of the user is challenging and requires mechanisms such as full-body tracking \cite{xu2018monoperfcap} or eye tracking \cite{zhang2017mpiigaze}.

Another solution for reducing the required data rate is to apply data compression.
For a 2D real-time video, current media codecs can achieve a compression ratios from 250:1 to 1,000:1 \cite{selinis2020internet, Ericsson2022}.
Similarly, format conversion and data compression can be applied to reduce the data size in holographic communication.
The authors in \cite{mekuria2016design} propose a lossy real-time color-encoding method by exploiting the inter-frame redundancy of point clouds.
Moreover, considering the strong correlation among different views in a hologram, multi-view coding (MVC) for LFV-based streaming is proposed in \cite{xiang2016big}, which improves the compression rate by analyzing both horizontal and vertical correlations of images in adjacent angles and tilts.
Meanwhile, many efforts have been made by standardization groups for the compression of holograms. 
For example, the Moving Picture Experts Group (MPEG) defined the video point cloud compression (V-PCC) by converting point clouds into two separate video sequences that capture the geometry and texture information, respectively \cite{schwarz2018emerging}.
The Joint Photographic Experts Group (JPEG) intended to provide a standard representation framework to facilitate the compression of LFV-based or point cloud-based content for holographic communication \cite{schelkens2019jpeg}.
Different codecs for hologram compression are evaluated in \cite{amirpour2021quality}, in which the authors study the compression and restructure of holograms.

Retransmissions due to data packet loss result in additional delay.
To avoid the retransmission delay, the lost data packets can be recovered based on predicted data according to historical information of an object such as its trajectory.
For example, packets can be recovered from an LSTM-based prediction of human actions and movements in 3D \cite{liu2016spatio} or a short-term prediction by analyzing the actions, movements, or gestures of users \cite{manolova2021context}.
By predicting content, data packets can be generated at the receiver side in the event of packet loss to reduce the delay in holographic communication \cite{strinati20196g}.

\subsubsection{Communication and Networking Solutions}
In addition to data processing, some communication and networking solutions have been investigated for satisfying delay and data rate requirements of holographic communication, including computing architectures, transport protocols, and physical layer technologies. 

In holographic communication, data captured from different sensors needs to be processed to form a 3D representation of the object, which is then rendered and reconstructed at the receiver side \cite{Javidi05}. 
However, the limited computing capability of local devices may lead to a long processing delay due to the high workload of data fusion and rendering \cite{hu2017survey}. 
Cloud computing is introduced to support high computing workloads for data processing in holographic communication. 
However, transmitting massive data to the cloud may result in a high communication delay \cite{wang2022task}, which is not suitable for real-time holographic communication.
One promising solution is to offload computing tasks to MEC servers for data processing, since MEC servers possess considerable computing capability and are placed close to users \cite{gupta20216g}. 
Thanks to network function virtualization (NFV), functions such as data fusion, data compression, and data rendering can be virtualized and flexibly deployed on MEC servers.
In this case, captured data from different sources can be aggregated, fused, and synchronized at an MEC server before rendering \cite{qian2022remote}.
Moreover, a multi-tier computing scheme is proposed for 6G networks, which can be utilized for holographic communication by integrating computing resources at cloud servers, MEC servers, and local devices, to achieve a low delay for data transmission and high computing capacity for data processing with collaboration among different servers \cite{yang2018meets, wang2022task}.
By integrating computing resources on different tiers, content can be processed at different servers to effectively utilize computing resources, and flexible computing resource management should be developed to facilitate multi-tier computing for holographic communication.
For example, split rendering is introduced for an MEC server and a local device to cooperatively decode and render holograms according to the content \cite{Ericsson2022}.

To satisfy the stringent delay and high reliability requirements of holographic communication, transport layer optimizations are also crucial. 
Current transport protocols, such as transmission control protocol (TCP) and user datagram protocol (UDP), can hardly satisfy the requirements of holographic communication.
To improve the reliability and delay performance in real-time communication, new protocols based on UDP are introduced, such as QUIC over HTTP/3 \cite{seufert2019quicker}.
Currently, the research on QUIC mainly focuses on traditional 2D video streaming services, while QUIC can serve as a potential solution for holographic communication, providing a quality-managed low-delay streaming option \cite{clemm2020toward}.
Moreover, the transmission of a hologram may consist of multiple substreams corresponding to different viewpoints, while the QoS requirement and the priority of each substream may be different.
In this case, the transmission of the most essential substreams needs to be prioritized.
To achieve this target, a new transport protocol is designed in \cite{rozen2021prism} for holographic communication to satisfy different QoS requirements of different flows by providing flow-level granular control.
In addition, an adaptive retransmission mechanism based on TCP is designed to reduce retransmissions by analyzing and differentiating packets \cite{clemm2020toward}. 
For example, only important data, such as the data used for rendering the part of the hologram in the center of the user's FoV, will be retransmitted if the related packets are lost, to reduce retransmissions. 

Finally, physical layer technologies are important to supporting a high data rate for holographic communication. 
In order to transmit high-resolution LFV-based holograms, holographic communication requires a data rate of several Tbps, while current 5G networks cannot support it \cite{shahraki2021comprehensive, david20186g}.
Featuring higher frequency and larger bandwidth compared with mmWave in 5G, THz communications have the potential to support holographic communication with Tbps-level data rate \cite{elayan2019terahertz, chen2019survey}.
To overcome the severe propagation loss of THz communication, dense deployment of access points and extremely narrow beams can be adopted to improve connection density and communication reliability \cite{8766143}.
Considering the absorption and reflection properties in the THz regime \cite{aazhang2019key}, the deployment of the THz base stations and the prediction of user motion require further investigation to provide sustainable LoS links for holographic communication \cite{chaccour2022seven}.
In addition to THz communications, visible light communication (VLC) can provide an alternative solution for holographic communication by providing large available bandwidth \cite{beysens2021blendvlc}. Featuring a high transmission data rate \cite{strinati20196g} and accurate positioning \cite{li2015human}, VLC can potentially support holographic communication as well as user tracking in an indoor environment. The coordination of THz communication and VLC is studied in \cite{wang2022meta} for providing a reliable service with a high data rate.

Table~\ref{Table2} provides a summary of the solutions discussed in this section as well as their limitations or costs.

\begin{table*}[hbt!]
    \caption{Potential solutions to immersive communications and the costs}\label{Table2}
    \renewcommand{\arraystretch}{1.5}%
    \begin{tabular}{p{2cm}|p{2.3cm}|p{7.5cm}|p{4cm}}
        \hline\hline
            \textbf{Use Cases} & \textbf{Objective} & \textbf{Solutions} & \textbf{Cost/Limitations} \\ 
        \hline\hline
          \multirow{7}*{\textbf{XR}} & {QoE maximization} & Balancing the robustness of viewpoint prediction error and video quality by optimization \cite{9939105} and machine learning \cite{9614984} methods & {Require centralized online management and learning} \\ \cline{2-4}
          ~& {Delay satisfaction improvement} & Viewpoint prediction \cite{10.1145/3229625.3229629}, communication protocol design \cite{yen2019streaming}, and adaptive computing offloading \cite{9246260} & Require user state characterization and data management \\ \cline{2-4}
            ~ & Computing process optimization & Edge caching \cite{Yaping} and multi-server and/or multi-device coordination \cite{Siyi, xie2021q}  & Require storage resources and increase computing complexity\\ \cline{1-4}
          \multirow{7}*{\textbf{Haptics}} & Delay reduction & Grant-free multiple access \cite{ali2021urllc, gao2021mac} & Increase interference management complexity \\ \cline{2-4}
          ~& Reliability enhancement & Diversity techniques \cite{tarneberg2017utilizing} and haptic data prediction \cite{maier2019towards,hou2019prediction} & {Increase resource consumption and computing complexity, respectively} \\ \cline{2-4}
            ~ & Data rate reduction & Perceptual haptic reduction \cite{steinbach2018haptic,holland2019ieee} & Require adaptive human perception modeling\\ \cline{1-4}
          \multirow{7}*{\textbf{Holography}} & Delay reduction & \multirow{2}*{Multi-tier computing \cite{Ericsson2022}} & Require collaboration of multiple servers \\ \cline{2-4}
          ~& Reliability enhancement & New designed transport protocol \cite{rozen2021prism}  & Require abundant data for fine-granular flow control \\ \cline{2-4}
            ~ & Data rate reduction & Content selection and compression \cite{clemm2020toward, mekuria2016design}& {Increase computing complexity}\\ \cline{2-4}
             ~ & High data rate support & THz communication \cite{chaccour2022seven} and VLC \cite{beysens2021blendvlc} & Limited communication range\\ \cline{3-4}
        \hline\hline
    \end{tabular}
\end{table*}

\section{Immersive Communication: Open Issues and Future Directions}\label{s:FDs}

Despite an increasing amount of studies and solutions for supporting XR, haptic communication, and holographic communication, there exist many open issues to address before immersive communications can popularize. To name a few, synchronization of multi-modal communications, user QoE modeling and enhancement, and intelligent network management for immersive communications remain to be  challenging problems. In this section, we present some major open issues in immersive communications and potential future directions to address these issues.

\subsection{Multi-Modal Communications}

While immersive communications have the potential to enhance user engagement and facilitate immersive interactions, effective network resource management for ensuring synchronized multi-modal perception in highly dynamic network environments is an open issue. The synchronization of multi-modal perception consists of two aspects: inter-stream (cross-modal) and intra-stream. First, the transmission of auditory, visual, and haptic data results in multiple data streams that should be synchronized in order to prevent motion sickness. For example, the time interval between perceived visual and tactile movement should not exceed 1 ms~\cite{van2017challenges}. Second, to enhance the immersive experience, a data stream can include multiple data substreams corresponding to different sensations, e.g., temperature and pressure, which also need synchronization. Data substreams corresponding to different DoF of an HI should be synchronized to maintain the stable perception of simultaneity, and data substreams transmitted from LIDAR sensors placed at different locations should be synchronized to render a 3D hologram precisely. There are many works that enable either intra-stream or inter-stream synchronization from the perspective of a single network layer~\cite{zhang2018towards,cizmeci2017multiplexing}. However, in order to synchronize multi-modal perception, both network-related and application-related information are necessary. This is because network resource management for multi-modal communications is affected by not only different data packet formats, data traffic patterns, and QoS requirements, but also different sensitivities of human perception. The cross-layer design of network protocols for multi-modal communications, which can support information sharing among different layers for efficient use of network resources, is a potential solution~\cite{she2020tutorial,kumar2018multi}. A higher-layer approach synchronizing multi-modal information can benefit from information on network conditions at lower layers, e.g., adaptively changing the priority of modalities in transport-layer multiplexing according to real-time physical-layer data rates. In addition, lower-layer approaches can take into account application-related information for efficient network resource management, e.g., timely adjusting the amount of radio resources allocated to a user in response to the dynamic sensitivity of the user's perception. \bl{Since multi-modal perception data in immersive communications can include personal biometric information of individual users, privacy challenges can arise in the transmission and processing of such data, such as biometric data leakage or profiling~\cite{shen2021data}.}

\subsection{AI-Native Immersive Communications}

AI techniques have demonstrated outstanding performance in identifying data correlations and analyzing device dynamics. As a result, some application functions using AI techniques, i.e., AI-enabled functions, have been developed for exploring unknown device states in immersive communications, such as viewpoint predictions in VR devices and haptic data prediction \cite{9749222}. To support increased service demands on immersive communications in 6G, AI-enabled functions will be deployed at network servers, i.e., cloud and edge servers \cite{9586568}. Accordingly, the network should support the entire lifecycle of AI for the functions, including data collection, data pre-processing, AI model training, inference, and AI model evaluation. By taking AI-enabled functions as the built-in component for supporting immersive communications, several potential future research directions should be investigated. First, AI-enabled functions can be configured according to network management policies for supporting immersive communications. For example, in haptic communication, the prediction horizon, i.e., the time window for the predicted information, of tactile and kinesthetic information can be adjusted to adapt to real-time network transmission and computing delay, AI-based prediction accuracy, and service reliability requirements. Second, efficient data management schemes can be developed, in which low-signaling-overhead and grant-free network management can be achieved by sharing the data obtained from AI-enabled functions. For example, in VR video delivery, network controllers can use a viewpoint prediction model or results from a viewpoint prediction function and allocate sufficient downlink communication resources to users with highly dynamic viewpoint movements. Additionally, effective resource management solutions should be developed to support AI model training in real-time, so that AI-enabled functions can be updated according to user behavior dynamics, where sufficient network resources should be allocated for supporting data collection and processing at edge and cloud servers. When supporting AI-native immersive communications, essential security issues should be addressed. For example, data and model poisoning attacks can lead to biased or incorrect results by injecting false samples into the training datasets and updating crafted local AI models in federated learning, respectively~\cite{khisamova2019artificial}.

\subsection{Time-Sensitive and Deterministic Networking}

The existing solutions mentioned in Section~\ref{s:Solus} can help reduce transmission delay in immersive communications. However, satisfying the stringent delay and reliability requirements of XR, haptic communication, and holographic communication, especially ms-level end-to-end delay, remains a challenge. Fortunately, the ongoing efforts of 3GPP, IEEE, and IETF in supporting time-sensitive networking (TSN) and deterministic networking (DetNet)~\cite{8412459, 8458130} provide solutions to meet the requirements of immersive communications~\cite{9855453}. The current efforts largely focus on the link and network layers (i.e., layers 2 and 3) and mostly target industrial networks~\cite{9855453}. Therefore, the corresponding solutions may not be readily applicable to all use cases of immersive communications. Potential future directions of TSN and DetNet for immersive communications include the followings. First, a comprehensive solution integrating existing TSN and DetNet designs for delay minimization can be important to immersive communications. For example, the joint design of coordinated sensing/capturing and communication (on the physical layer), traffic shaping and scheduling (on the link layer), flow identification and packet treatment (on the network layer), and viewpoint/haptic data prediction (on the application layer) can help reduce the end-to-end delay in immersive communications. Second, instead of treating different data streams in a mutli-modal communication separately, joint prioritization and resource orchestration for different types of data given their respective delay and jitter requirements is another promising direction. Third, integrating environment-aware and service-oriented network management paradigms can potentially enable TSN and DetNet for immersive communications. An example is to incorporate adaptive radio access network (RAN) function splitting, network slicing, and AI-driven network management to minimize delay and jitter by customizing for a specific service and adapting to the network environment.

\subsection{QoE-Oriented Networking}

While QoS provisioning from a network perspective benefits the transmission of XR content, haptic information, and holograms, as detailed in Section~\ref{s:Solus}, evaluating and guaranteeing individual users' QoE is crucial in providing them an immersive experience. This is because many factors, besides communication network conditions, can affect user experience in immersive communications, including coding, compression, and human perception. Therefore, QoE-oriented networking from users' perspective is a promising network management paradigm to support immersive communications in the 6G era, including two potential aspects: personalized QoE modeling and QoE-oriented network resource management. First, existing works on immersive communications have limitations on personalizing QoE models for individual users. Conventional QoE modeling are based on either subjective tests or objective quality assessments~\cite{tasaka2022empirical}. The former, conducted in relatively static laboratory environments, is costly and inapplicable in dynamic network environments, whereas the latter, evaluated by empirical human perception models, does not differentiate individual users~\cite{barakabitze2019qoe,ruan2021survey}. Finding a way to model personalized QoE while adapting to dynamic network environments remains an open issue. Second, managing network resources to guarantee the QoE of individual users in immersive communications necessitates user-level information. Even if several users request the same service, they may have different resource demands for improving their  
QoE~\cite{kougioumtzidis2022qoe}. For example, due to the difference in the sensitivity of haptic perception, e.g., reaction time, the haptic sensors of interest and the scan time for each haptic sensor may differ in supporting different users, yielding different communication and computing resource demands~\cite{coutinho2022design}. In the 6G era, the paradigm of digital twins can be a potential solution for QoE-oriented networking. Specifically, individual users can be characterized by creating user digital twins, including user data profiles that contain extensive well-organized user data, and a variety of digital twin functions that support flexible and customized data collection and analysis~\cite{shen2021holistic}. Both personalized QoE modeling and QoE-oriented network resource management for immersive communications can benefit from extensive timely updated and fine-grained user-level information~\cite{wang2021interactive}. \bl{Although QoE-centric networking can provide users with immersive experiences based on the preferences and features of individual users, privacy issues, such as unconscionable behavioral profiling and improper uses of the profiles, should be addressed when collecting and processing data with user preference information~\cite{nguyen2021security}.}

\subsection{\bl{Collaborative Multi-tier Computing}}
Research on multi-tier computing is still at a nascent stage~\cite{yang2019multi}. In the 6G era, collaborative multi-tier computing can be a promising computing paradigm by leveraging the various characteristics of computing servers, such as service coverage and resource capacity. There are two research directions to facilitate immersive communications. First, computing tasks corresponding to different steps of immersive communications can be executed on different computing servers. Different steps of immersive communications may have different network resource demands, e.g., I/O-intensive data fusion tasks and CPU-intensive data encoding tasks require different communication and computing resources~\cite{gao2021cufsdaf}. Selecting proper computing servers for each step based on the features of computing servers and the resource demands of the step is beneficial for satisfying stringent QoS requirements of immersive communications. Second, context data management across computing servers at different tiers plays an important role in supporting immersive communications. A significant percentage of computing tasks in immersive communications will be stateful, meaning that context data are required during task execution, e.g., volumetric media objects or holograms in rendering~\cite{zhou2022digital,gao2022improve}. When stateful computing tasks are executed on a computing server, the required context data should be either stored locally on the computing server or downloaded remotely from other computing servers. As a result, managing context data, e.g., selecting proper computing servers from different tiers to proactively store context data based on the computing task arrival and mobility patterns of individual users, will have a significant impact on the performance of immersive communications. While collaborative multi-tier computing provides more options for context data management than conventional MEC, the coordination of computing servers at different tiers can significantly complicate the problem of context data management. 
In addition, establishing reliable trust relationships between computing servers and among computing servers and users, as well as measuring the credibility of users, is an open and important research direction in collaborative multi-tier computing for immersive communications~\cite{shen2022blockchain}.

\subsection{New Network Architecture}

Network architecture innovation is indispensable for a widespread realization of immersive communications, and innovations building on recent developments for 6G architecture are promising future directions. The need for new architectures manifests in several aspects. First, the computing-intensive nature of immersive communications, rooted from processing and compressing 3D data, predicting viewpoints and haptics data, and reconstructing 3D objects, demands a network architecture with extensive computing resources and reliable computing service provisioning. As a result, a heterogeneous network with multi-tier computing architecture~\cite{yang2019multi,zhou2022digital}, featuring on-demand and collaborative computing task offloading and scheduling across the network, is important to immersive communications yet open to investigation at the moment. Second, as networks become increasingly complex and the requirements of immersive communications become exceedingly stringent, supporting immersive communications in 6G requires a network architecture with unprecedented scalability, flexibility, and adaptivity. A 6G architecture integrating digital twins, network slicing, and pervasive AI~\cite{shen2021holistic} can be a foundation to immersive communications. Third, considering the diverse delay requirements of different XR, haptic communication, and holographic communication use cases, the Open-RAN (O-RAN) architecture featuring realtime, near-realtime, and non-realtime layers can benefit service differentiation in RAN management for immersive communications~\cite{abdalla2022toward}. Last, considering different user preferences and diverse user devices, a new architecture enabling user-centric networking, such as the everyone-centric architecture in~\cite{9839652}, has a potential to empower immersive communications. However, as none of the above architectures is developed specifically for immersive communications, new designs and customizations based on them for supporting {immersive communications} are open for investigation.


\section{Conclusion}\label{s:Con}

In this article, we have delved into immersive communications towards 6G and presented a comprehensive review of the related concepts, representative use cases, technical challenges and potential solutions, and future directions. Focusing on XR, haptic communication, and holographic communication, we have illustrated their general procedures, network requirements, and recent developments in the context of a vision for 6G. Despite abundant emerging use cases and exciting recent advancements, we have shown that many challenges are yet to be conquered before the envisioned prosperity of immersive communications can occur. In particular, the exceeding transmission rate, delay, and reliability requirements, further complicated by the multi-modal and computing-intensive features of immersive communications, indicate the necessity of an unprecedented amount of communication and computing resources as well as novel paradigms such as AI-native communication, multi-tier computing, and user-centric networking. 

To respond to the challenges posed by supporting immersive communications and promote further research, we have presented various solutions and future directions in this survey. From physical-layer technologies such as Terahertz communications to application-layer solutions such as user behavior prediction, advances in each layer will contribute to the realization of immersive communications. Meanwhile, new paradigms envisioned for 6G, such as QoE-oriented networking and AI-native communications, represent promising future directions for researchers in the field to explore.

The paradigm shift to immersive communications is truly exciting and inspiring, especially when viewed in the context of the evolution toward 6G. Many opportunities exist, and more will emerge for researchers and engineers in the fields of communications, networking, and computer science to realize immersive communications. We hope this review inspires further interest among fellow researchers and provides fundamental knowledge on related research, thereby contributing to this much-anticipated paradigm shift and making immersive communications the next reality.   




\section*{Acknowledgments}
The authors would like to thank Dr.~Dongxiao Liu for his helpful comments related to the security and privacy issues in immersive communications.

\bibliography{test1}
\bibliographystyle{IEEEtran}

\begin{IEEEbiography}[{\includegraphics[width=1in,height=1.25in,clip,keepaspectratio]{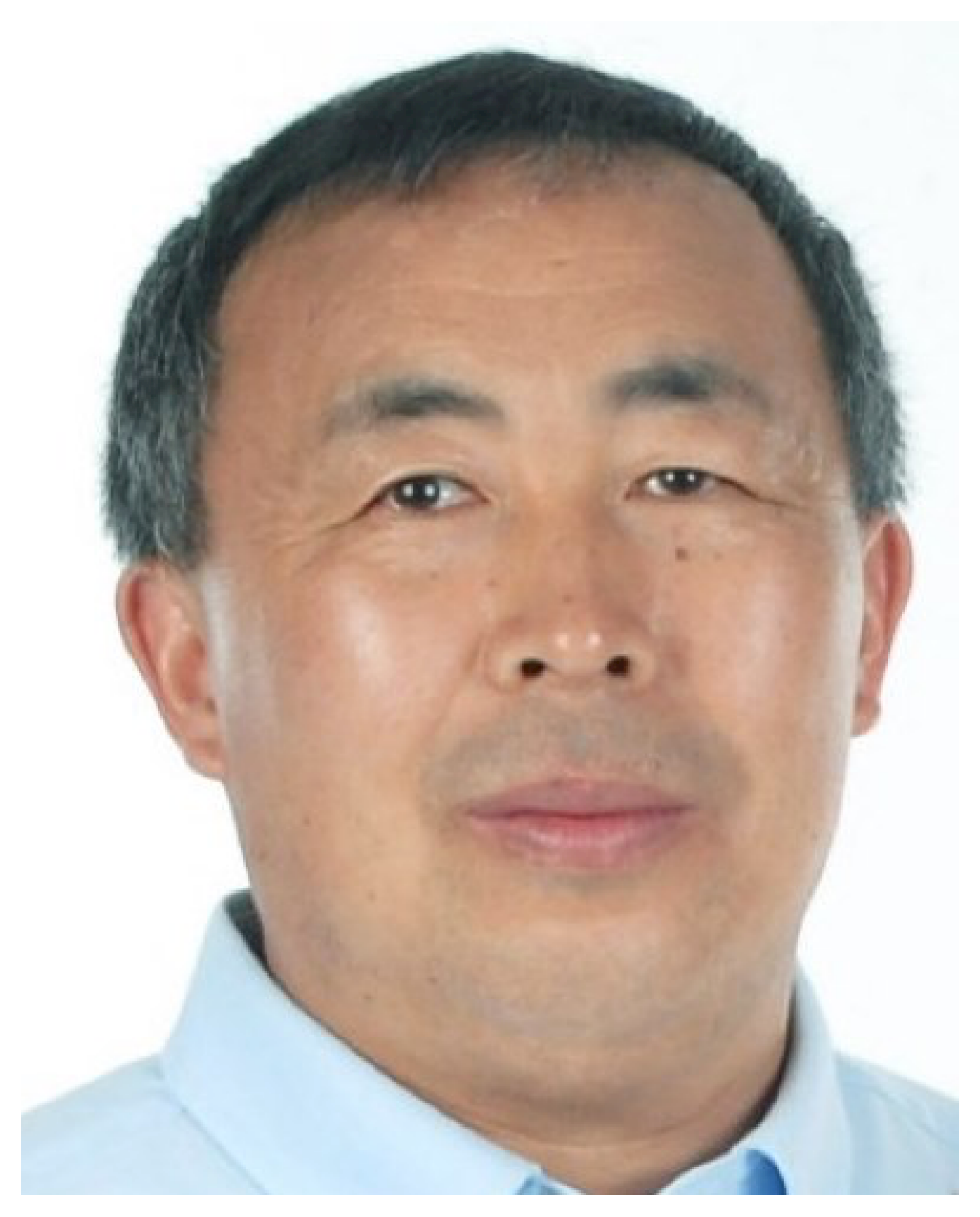}}]{Xuemin (Sherman) Shen}(M'97-SM'02-F'09) received the Ph.D. degree in electrical engineering from Rutgers University, New Brunswick, NJ, USA, in 1990. He is a University Professor with the Department of Electrical and Computer Engineering, University of Waterloo, Canada. His research focuses on network resource management, wireless network security, Internet of Things, 5G and beyond, and vehicular networks. Dr. Shen is a registered Professional Engineer of Ontario, Canada, an Engineering Institute of Canada Fellow, a Canadian Academy of Engineering Fellow, a Royal Society of Canada Fellow, a Chinese Academy of Engineering Foreign Member, and a Distinguished Lecturer of the IEEE Vehicular Technology Society and Communications Society.

Dr. Shen received the Canadian Award for Telecommunications Research from the Canadian Society of Information Theory (CSIT) in 2021, the R.A. Fessenden Award in 2019 from IEEE, Canada, Award of Merit from the Federation of Chinese Canadian Professionals (Ontario) in 2019, James Evans Avant Garde Award in 2018 from the IEEE Vehicular Technology Society, Joseph LoCicero Award in 2015 and Education Award in 2017 from the IEEE Communications Society, and Technical Recognition Award from Wireless Communications Technical Committee (2019) and AHSN Technical Committee (2013). He has also received the Excellent Graduate Supervision Award in 2006 from the University of Waterloo and the Premier’s Research Excellence Award (PREA) in 2003 from the Province of Ontario, Canada. He served as the Technical Program Committee Chair/Co-Chair for IEEE Globecom’ 16, IEEE Infocom’14, IEEE VTC’10 Fall, IEEE Globecom’07, and the Chair for the IEEE Communications Society Technical Committee on Wireless Communications. Dr. Shen is the President of the IEEE Communications Society. He was the Vice President for Technical \& Educational Activities, Vice President for Publications, Member-at-Large on the Board of Governors, Chair of the Distinguished Lecturer Selection Committee, Member of IEEE Fellow Selection Committee of the ComSoc. Dr. Shen served as the Editor-inChief of the IEEE IoT JOURNAL, IEEE Network, and IET Communications.
\end{IEEEbiography}

\begin{IEEEbiography}[{\includegraphics[width=1in,height=1.25in,clip,keepaspectratio]{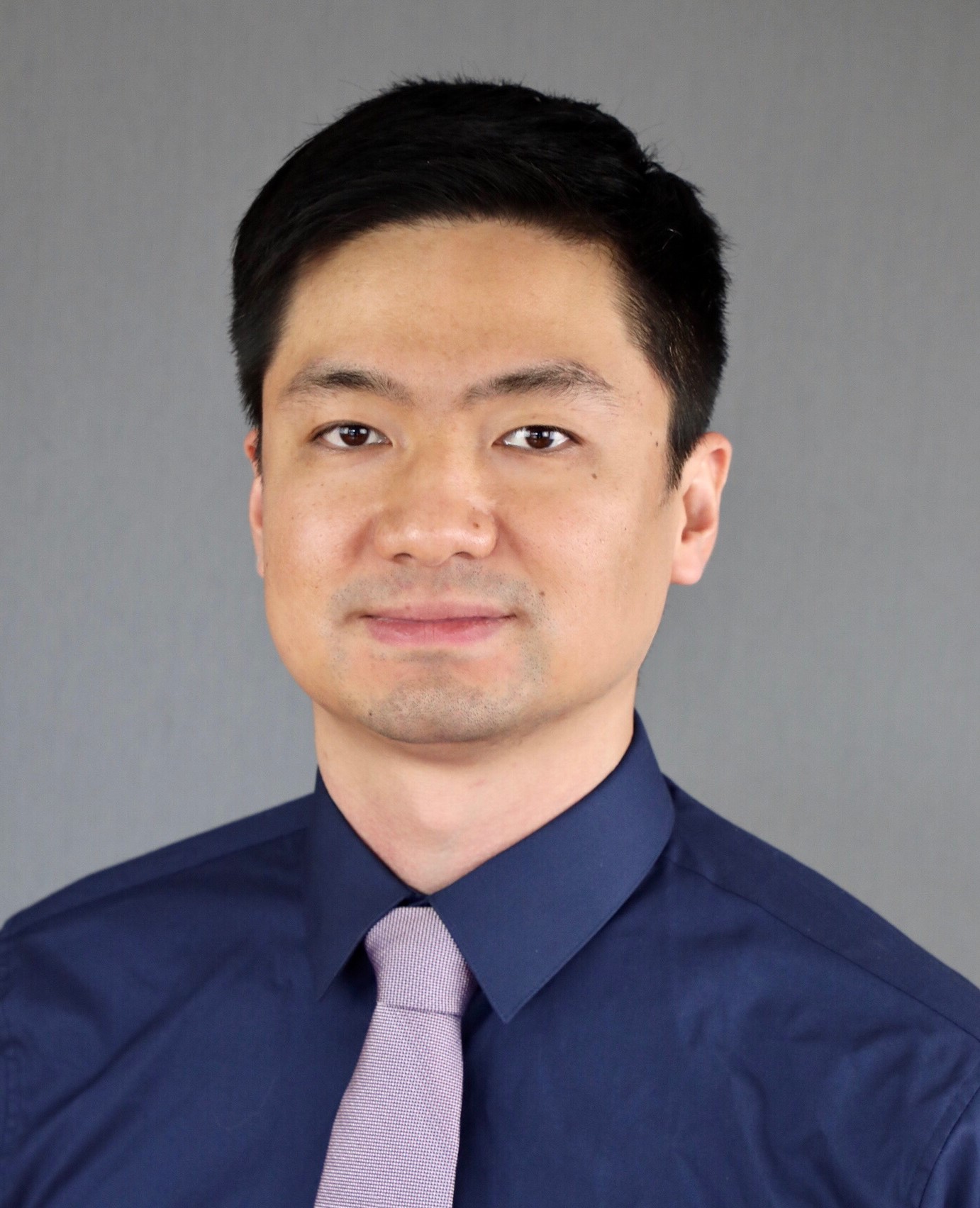}}]{Jie Gao} (S’13-M’17-SM’21) received the B.Eng. degree in electronics and information engineering from the Huazhong University of Science and Technology, Wuhan, China, in 2007, and the M.Sc. and Ph.D. degrees in electrical engineering from the University of Alberta, Edmonton, AB, Canada, in 2009 and 2014, respectively. He was a postdoctoral fellow with Toronto Metropolitan University, Toronto, ON, Canada, from 2017 to 2019, a research associate with the University of Waterloo, Waterloo, ON, Canada, from 2019 to 2020, and an assistant professor with the Department of Electrical and Computer Engineering, Marquette University, Milwaukee, WI, USA, from 2020 to 2022. Dr. Gao is currently an assistant professor in the School of Information Technology, Carleton University. His research interests include machine learning for communications and networking, network virtualization and digital twins, Internet of Things (IoT) and industrial IoT solutions, and next-generation wireless networks in general. 
\end{IEEEbiography}

\begin{IEEEbiography}[{\includegraphics[width=1in,height=1.25in,clip,keepaspectratio]{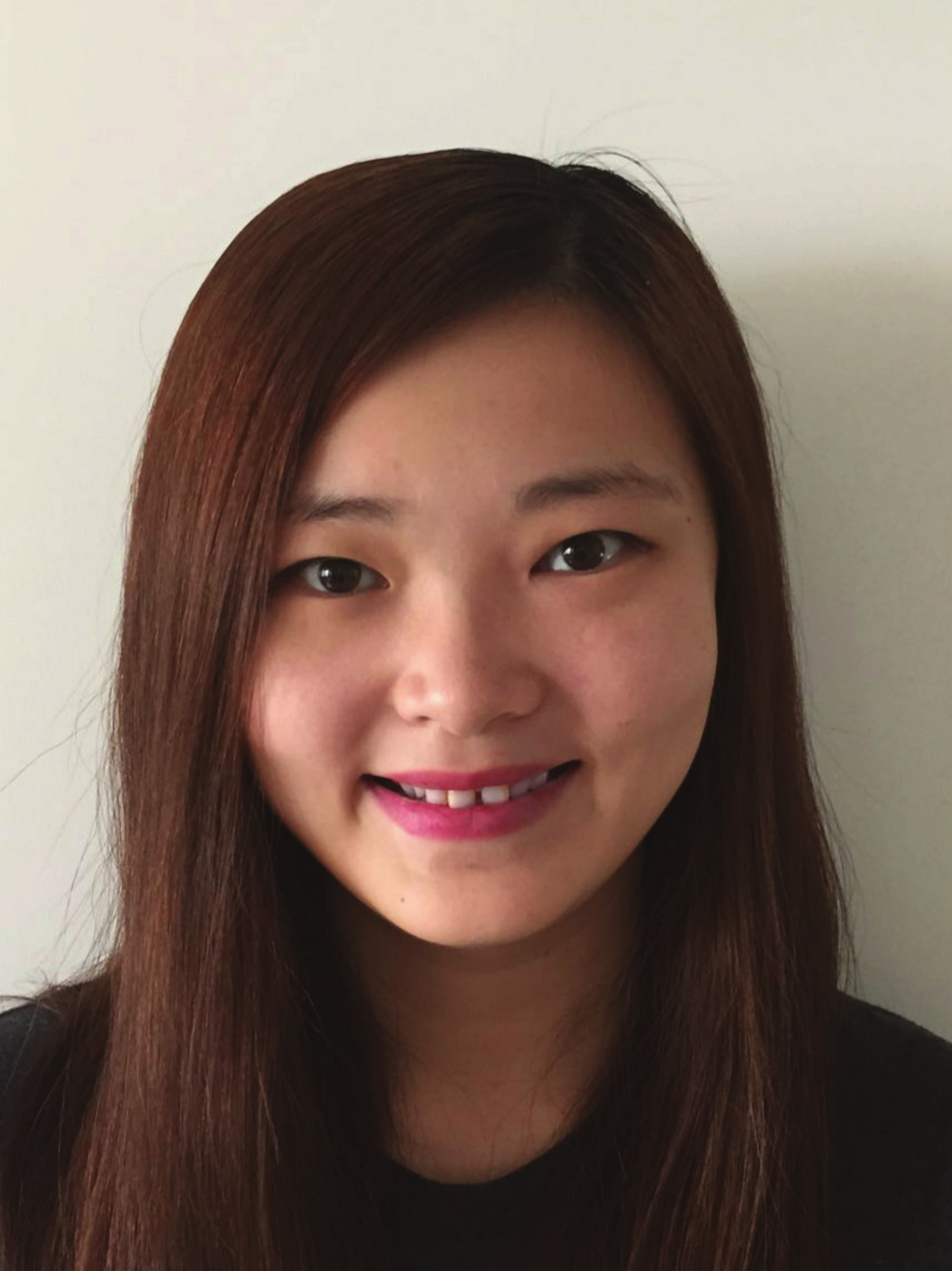}}]{Mushu Li}(S'18-M'21) 
    received the Ph.D. degree in Electrical and Computer Engineering from the University of Waterloo, Canada, in 2021. She received the B.Eng. degree from the University of Ontario Institute of Technology (UOIT), Canada, in 2015, the M.A.Sc. degree from Ryerson University, Canada, in 2017. She is currently a Postdoctoral Fellow at Toronto Metropolitan University, ON, Canada. She was a Postdoctoral Fellow with the University of Waterloo, ON, Canada, from 2021 to 2022. Her research interests include mobile edge computing, the system optimization in wireless networks, and machine learning-assisted network management. She was the recipient of Natural Science and Engineering Research Council of Canada (NSERC) Postdoctoral Fellowship (2022), NSERC Canada Graduate Scholarship (2018), and Ontario Graduate Scholarship (2015 and 2016). 
\end{IEEEbiography}

\begin{IEEEbiography}[{\includegraphics[width=1in,height=1.25in,clip,keepaspectratio]{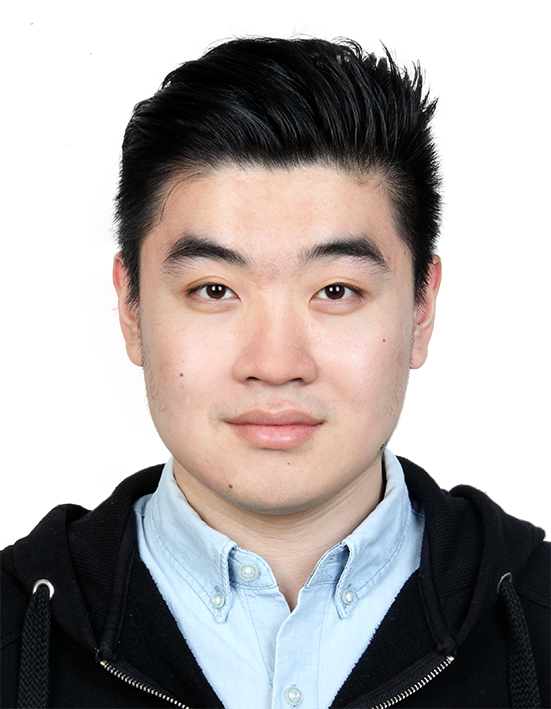}}]{Conghao Zhou}
	(S'19-M'22) received the Ph.D. degree in Electrical and Computer Engineering from University of Waterloo, Waterloo, ON, Canada, in 2022. He received the B.Eng. degree from Northeastern University, Shenyang, China, in 2017 and the M.Sc. degree from University of Illinois at Chicago, Chicago, IL, USA, in 2018. He is currently a Postdoctoral Fellow at University of Waterloo, Waterloo, ON, Canada. His research interests include space-air-ground integrated networks, multi-tier computing, and machine learning in wireless networks.
\end{IEEEbiography}

\begin{IEEEbiography}[{\includegraphics[width=1in,height=1.25in,clip,keepaspectratio]{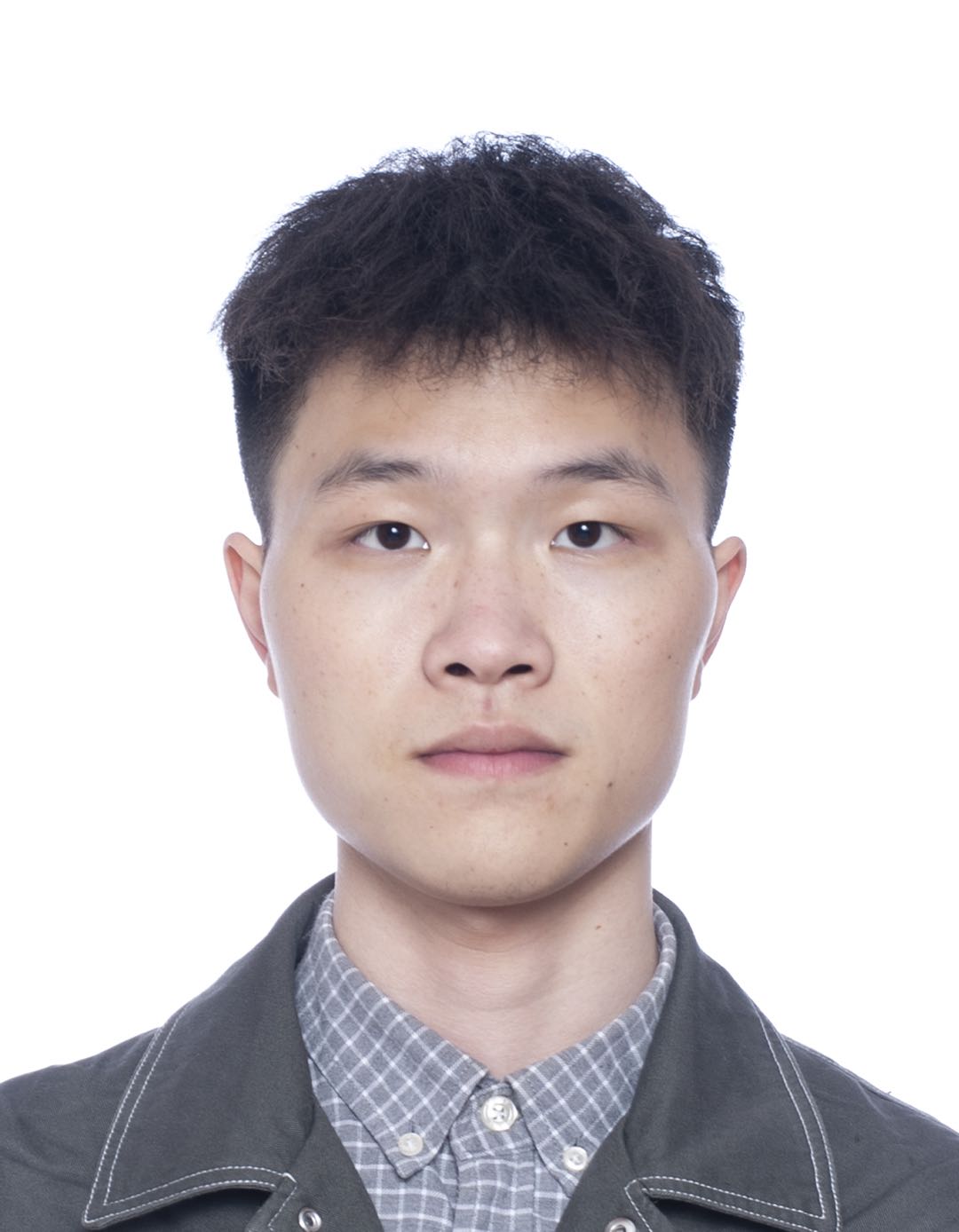}}]{Shisheng Hu}
	(S'19) received the B.Eng. degree and the M.A.Sc. degree from University of Electronic Science and Technology of China (UESTC), Chengdu, China, in 2018 and 2021, respectively. He is currently working toward the Ph.D. degree with the Department of Electrical and Computer Engineering, University of Waterloo, Waterloo, ON, Canada. His research interests include AI for wireless networks and networking for AI.
\end{IEEEbiography}

\begin{IEEEbiography}[{\includegraphics[width=1in,height=1.25in,clip,keepaspectratio]{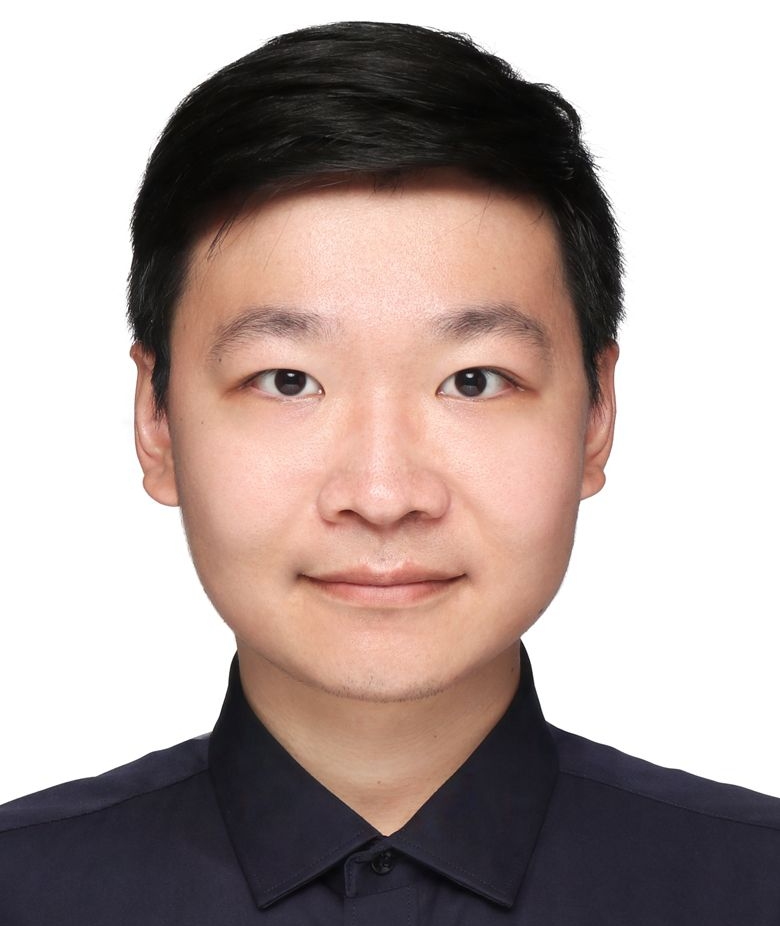}}]{Mingcheng He}
	(S'21) received the B.S. and M.Eng. degrees from Shanghai Jiao Tong University, Shanghai, China, in 2017 and 2020, respectively. He is currently working toward the Ph.D. degree with the Department of Electrical and Computer Engineering, University of Waterloo, Waterloo, ON, Canada. His research interests include satellite-terrestrial integration networks and machine learning in wireless networks.
\end{IEEEbiography}

\begin{IEEEbiography}[{\includegraphics[width=1in,height=1.25in,clip,keepaspectratio]{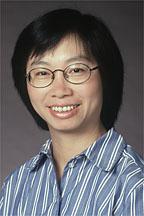}}]{Weihua Zhuang}(M'93-SM'01-F'08) received the B.Sc. and M.Sc. degrees from Dalian Marine University, China, and the Ph.D. degree from the University of New Brunswick, Canada, all in electrical engineering. She is a University Professor and a Tier I Canada Research Chair in Wireless Communication Networks at University of Waterloo, Canada. Her research focuses on network architecture, algorithms and protocols, and service provisioning in future communication systems. She is the recipient of 2021 Women’s Distinguished Career Award from IEEE Vehicular Technology Society, 2021 Technical Contribution Award in Cognitive Networks from IEEE Communications Society, 2021 R.A. Fessenden Award from IEEE Canada, and 2021 Award of Merit from the Federation of Chinese Canadian Professionals in Ontario. She was the Editor-in-Chief of the IEEE Transactions on Vehicular Technology from 2007 to 2013, General Co-Chair of 2021 IEEE/CIC International Conference on Communications in China (ICCC), Technical Program Chair/Co-Chair of 2017/2016 IEEE VTC Fall, Technical Program Symposia Chair of 2011 IEEE Globecom, and an IEEE Communications Society Distinguished Lecturer from 2008 to 2011. She is an elected member of the Board of Governors and the Executive Vice President of the IEEE Vehicular Technology Society. Dr. Zhuang is a Fellow of the IEEE, Royal Society of Canada, Canadian Academy of Engineering, and Engineering Institute of Canada.
\end{IEEEbiography}

\end{document}


\onecolumn
\firstpage{1}

\title[Towards Immersive Communications in 6G]{{\helveticaitalic{Towards Immersive Communications in 6G}}}

\maketitle

\abstract{

}

\section{Introduction}\label{s:Intro}

\section{Use Cases}\label{s:UCs}

\section{Components of Immersive Communications}\label{s:Comps}

\subsection{Concepts}\label{ss:Concs}

\subsubsection{AR/VR/MR/XR}\label{sss:XR}

\subsubsection{Haptics and Tactile Networks}\label{sss:HapTac}

\subsubsection{Holographic Communications}\label{sss:Holo}

\subsection{Procedures}\label{s:Procs}

\subsection{Requirements}\label{s:Reqs}

\subsubsection{Bandwidth/Throughput}\label{sss:Bandw}

\subsubsection{Delay}\label{sss:delay}

\subsubsection{Reliability}\label{sss:Reli}

\section{Solutions}\label{s:Solus}

\section{Future Directions}\label{s:FDs}

For more information on Supplementary Material and for details on the different file types accepted, please see \href{http://home.frontiersin.org/about/author-guidelines#SupplementaryMaterial}{the Supplementary Material section} of the Author Guidelines.

Figures, tables, and images will be published under a Creative Commons CC-BY licence and permission must be obtained for use of copyrighted material from other sources (including re-published/adapted/modified/partial figures and images from the internet). It is the responsibility of the authors to acquire the licenses, to follow any citation instructions requested by third-party rights holders, and cover any supplementary charges.


\subsection{Figures}


\begin{figure}[htbp]
\begin{center}
\includegraphics[width=9cm]{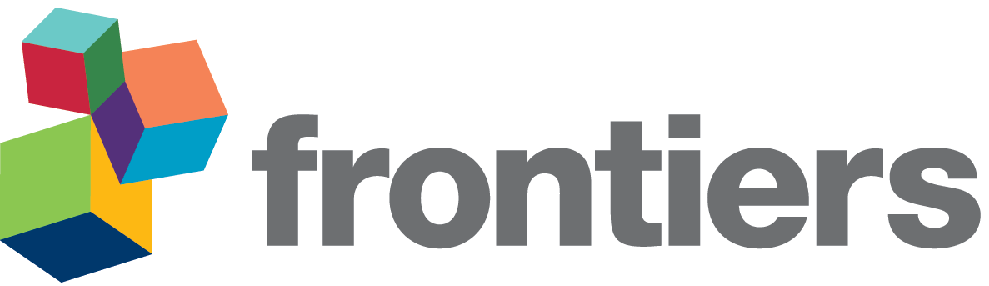}
\end{center}
\caption{ Enter the caption for your figure here.  Repeat as  necessary for each of your figures}\label{fig:1}
\end{figure}

\setcounter{figure}{2}
\setcounter{subfigure}{0}
\begin{subfigure}
\setcounter{figure}{2}
\setcounter{subfigure}{0}
    \centering
    \begin{minipage}[b]{0.5\textwidth}
        \includegraphics[width=\linewidth]{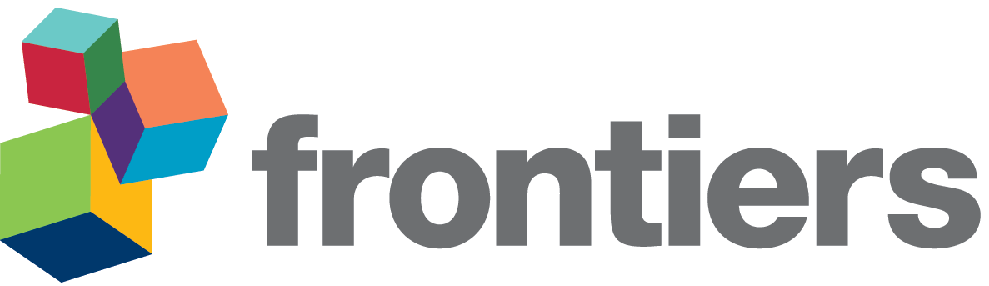}
        \caption{This is Subfigure 1.}
        \label{fig:Subfigure 1}
    \end{minipage}  
   
\setcounter{figure}{2}
\setcounter{subfigure}{1}
    \begin{minipage}[b]{0.5\textwidth}
        \includegraphics[width=\linewidth]{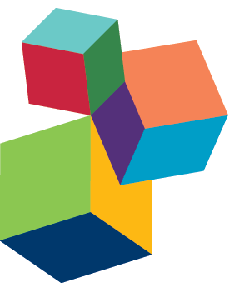}
        \caption{This is Subfigure 2.}
        \label{fig:Subfigure 2}
    \end{minipage}

\setcounter{figure}{2}
\setcounter{subfigure}{-1}
    \caption{Enter the caption for your subfigure here. \textbf{(A)} This is the caption for Subfigure 1. \textbf{(B)} This is the caption for Subfigure 2.}
    \label{fig: subfigures}
\end{subfigure}


